\documentclass[a4wide]{article}
\usepackage[margin=1.2in]{geometry}

\usepackage{type1cm}        
\usepackage{makeidx}         
\usepackage{graphicx}        
\usepackage{multicol}        
\usepackage[bottom]{footmisc}

\usepackage{newtxtext}       %
\usepackage{newtxmath}       

\makeindex             

\usepackage[utf8]{inputenc}
\usepackage[american]{babel}

\makeatletter 
\def\ifEmpty#1{\def\@temp{#1}\ifx\@temp\@empty} 
\makeatother
\newcommand{\lyapack}{\textsf{LyaPack}}
\newcommand{\mess}{\textsf{M.E.S.S.}}
\newcommand{\mmess}{\mbox{\textsf{M-}\mess{}}}
\newcounter{mymac@matlab}
\newcommand{\matlab}{MATLAB%
  \ifnum\value{mymac@matlab}<1%
  \textsuperscript{\textregistered}%
  \setcounter{mymac@matlab}{1}%
  \fi%
}
\newcommand{\octave}{\textsf{GNU Octave}}
\newcommand{\intel}{Intel\textsuperscript{\textregistered}}
\newcommand{\tran}{\ensuremath{\mkern-1.5mu\mathsf{T}}}
\newcommand\N{\ensuremath{\mathbb{N}}}
\newcommand\R{\ensuremath{\mathbb{R}}}
\newcommand{\Rn}{\ensuremath{\R^{n}}}
\newcommand{\Rnn}{\ensuremath{\R^{n\times n}}}
\newcommand{\Rpn}{\ensuremath{\R^{p\times n}}}
\newcommand{\Rnm}{\ensuremath{\R^{n\times m}}}
\newcommand{\Rpm}{\ensuremath{\R^{p\times m}}}
\newcommand{\Rrm}{\ensuremath{\R^{r\times m}}}
\newcommand{\Rnr}{\ensuremath{\R^{n\times r}}}
\newcommand{\Rpr}{\ensuremath{\R^{p\times r}}}
\newcommand{\Rrr}{\ensuremath{\R^{r\times r}}}
\newcommand{\nrm}    [2][] {\ensuremath{\left\lVert #2 \right\rVert\ifEmpty{#1}\else_{#1}\fi}}
\let\norm\nrm{}

\newcommand{\tA}{\ensuremath{\tilde{A}}}
\newcommand{\tB}{\ensuremath{\tilde{B}}}
\newcommand{\tC}{\ensuremath{\tilde{C}}}
\newcommand{\tE}{\ensuremath{\tilde{E}}}
\newcommand{\cH}{\ensuremath{\mathcal{H}}}
\newcommand{\hA}{\ensuremath{\hat{A}}}
\newcommand{\hB}{\ensuremath{\hat{B}}}
\newcommand{\hC}{\ensuremath{\hat{C}}}
\newcommand{\hE}{\ensuremath{\hat{E}}}
\newcommand{\hH}{\ensuremath{\hat{H}}}
\newcommand{\hx}{\ensuremath{\hat{x}}}
\newcommand{\hy}{\ensuremath{\hat{y}}}

\usepackage{tabularx}
\usepackage{ragged2e}
\newcolumntype{L}{>{\RaggedRight}X}

\usepackage{listings} 
\usepackage[hidelinks]{hyperref}
\usepackage{url}
\usepackage{doi}

\usepackage{tikz}
\usepackage{pgfplots} 
\pgfplotsset{compat = 1.13, 
  colormap name = viridis, 
  unbounded coords = jump, 
  every axis plot/.append style = {%
    line width = 1.5pt 
  } 
} 
\usepgfplotslibrary{external} 
\tikzset{external/system call = {%
    lualatex \tikzexternalcheckshellescape
    -halt-on-error
    -interaction=batchmode
    -jobname "\image" "\texsource"}} 
\tikzexternaldisable{} 

\usepackage{filemod}

\usepackage{caption} 
\usepackage{subcaption} 

\def\myvdots{\vbox{\baselineskip=4pt \lineskiplimit=0pt 
\kern3pt \hbox{.}\hbox{.}\hbox{.}}}
\def\myddots{\mathinner{\mkern1mu\raise7pt\vbox{\kern3.5pt\hbox{.}}\mkern2mu 
\raise4pt\hbox{.}\mkern2mu\raise1pt\hbox{.}\mkern1mu}}

\title{Matrix Equations, Sparse Solvers: \mmess{}-2.0.1 --- Philosophy, Features
  and Application for (Parametric) Model Order Reduction}
\author{Peter Benner\footnotemark[1] \and Martin Köhler\footnotemark[1] \and Jens Saak\thanks{Max Planck Institute for Dynamics of Complex Technical Systems,
  Sandtorstr. 1, 39106 Magdeburg, Germany,
  \texttt{\{benner,koehlerm,saak\}@mpi-magdeburg.mpg.de}}
}

\begin{document}
\maketitle
\abstract{Matrix equations are omnipresent in
  (numerical) linear algebra and systems theory. Especially in model order
  reduction (MOR) they play a key role in many balancing based reduction methods
  for linear dynamical systems. When these systems arise from spatial
  discretizations of evolutionary partial differential equations, their
  coefficient matrices are typically large and sparse. Moreover, the
  numbers of inputs and outputs of these systems are typically far smaller than
  the number of spatial degrees of freedom. Then, in many situations the
  solutions of the corresponding large-scale matrix equations are observed to
  have low (numerical) rank. This feature is exploited by \mmess{} to find
  successively larger low-rank factorizations approximating the solutions. This
  contribution describes the basic philosophy behind the implementation and the
  features of the package, as well as its application in the model order
  reduction of large-scale linear time-invariant (LTI) systems and parametric
  LTI systems.}

\section{Introduction}\label{sec:BKS20-introduction}
The \mmess{} toolbox~\cite{SaaKB-mmess-all-versions} for \matlab{} (or package
for \octave) in version~2.0.1 focuses on the solution of large-scale
symmetric algebraic and differential matrix equations and their application in
model order reduction (MOR) and linear-quadratic regulator (LQR) problems. The
basis for all considerations and problem formulations are linear dynamical
systems of the form
\begin{equation*}
  \begin{split}
    E\dot x(t) &= A x(t) + B u(t),\\
    y(t) &= C x(t) + D u(t),
  \end{split}\label{eq:BKS20-sys}\tag{\(\Sigma\)}
\end{equation*}
where \(E,A\in\Rnn\), \(B\in\Rnm\), \(C\in\Rpn\), \(D\in\Rpm\), and 
\(x(t)\in\Rn\), for all time instances \(t\in[0,T]\).
We assume that \(E\) is invertible, and often in addition
that~\eqref{eq:BKS20-sys} is asymptotically stable.

Some of the supported matrix equations have applications in
\(H_{\infty}\)-control, where the slightly more structured system
\begin{equation*}
  \begin{split}
    E\dot x(t) &= A x(t) + B_{1} u(t) + B_{2} w(t),\\
    y(t) &= C_{1} x(t) + D_{11} u(t) + D_{12} w(t),\\
    z(t) &= C_{2} x(t) + D_{21} u(t) + D_{22} w(t),
  \end{split}\label{eq:BKS20-sysinf}\tag{\(\Sigma_{\infty}\)}
\end{equation*}
is considered.

\mmess{} aims at systems, where \(n\in\N\) is too large to store an
\(n\times n\) matrix in the computer's memory. This will usually be accounted
for by the facts, that 
\(p,m\ll n\) and \(E\), \(A\) are sparse, or have a sparse realization that we
can exploit in computations. We present more details about the exploitable
structures in Section~\ref{sec:BKS20-philosophy_features}.

Similarly, for systems~\eqref{eq:BKS20-sysinf} the matrices \(B_{1}\),
\(B_{2}\), \(C_{1}\), \(C_{2}\) are considered thin and rectangular and the
parts \(D_{ij}\), \(i,j\in\{1,2\}\) correspondingly small.

The contribution of this document is two-fold. On the one hand, we give the
first concise introduction to \mmess{}, its general philosophy and current
features. On the other hand, we show how the software, that is in core intended
for the solution of large-scale matrix equations, can be employed in the
implementation of basic parameteric model order reduction (PMOR) methods for
systems of the form~\eqref{eq:BKS20-sys}.

Before moving on to the historical evolution of the package, we state the
equations that can currently be solved by \mmess{}. The following is a list of
all equations for which at least one solver function exists:
\begin{align*}
  \intertext{\bfseries Algebraic Lyapunov equations}
  \begin{split}
    0 &= APE^{\tran} + EPA^{\tran} + BB^{\tran}\\
    0 &= A^{\tran}QE + E^{\tran}QA + C^{\tran}C
  \end{split}\tag{CALE}\label{eq:BKS20-ALE}\\
  \intertext{\bfseries Algebraic Riccati equations}
  \begin{split}
    0 &= APE^{\tran} + EPA^{\tran} + BB^{\tran} - EPC^{\tran}C PE^{\tran}\\
    0 &= A^{\tran}QE + E^{\tran}QA + C^{\tran}C - E^{\tran}QBB^{\tran}QE    
  \end{split}\tag{CARE}\label{eq:BKS20-ARE}\\
  &\\
  \begin{split}
    0 &= \tA PE^{\tran} + EP\tA^{\tran} + \tB_{1}\tB_{1}^{\tran} -
    EP\left(\tC_{1}^{\tran}\tC_{1} - \tC_{2}^{\tran}\tC_{2}\right) PE^{\tran}\\
    0 &= \tA^{\tran}QE + E^{\tran}Q\tA + \tC_{1}^{\tran}\tC_{1} -
    E^{\tran}Q\left(\tB_{1}\tB_{1}^{\tran} - \tB_{2}\tB_{2}^{\tran}\right) QE
  \end{split}\tag{$\cH_{\infty}-ARE$}\label{eq:BKS20-iARE}
\end{align*}
In the last pair of equations, the matrix \(\tA\) is sparse plus low-rank
(splr), i.e.\ \(\tA = A + UV^{\tran}\), where \(U\), \(V\) are
tall and skinny.  Moreover the matrices \(\tB_{1}\), \(\tB_{2}\), \(\tC_{1}\),
\(\tC_{2}\)  are derived from the given system data by scaling
and potentially rotation of the matrices \(B_{1}\), \(B_{2}\), \(C_{1}\),
\(C_{2}\). 

For finite time horizon linear-quadratic control problems, one needs to solve
differential Riccati equations. We restrict to providing only the controller
equations here, while the dual ``filter-type'' equations are supported as well. 
\begin{align*}
  \intertext{\bfseries Autonomous differential Riccati equations}
  -E^{\tran}\dot{Q}(t)E &= A^{\tran}Q(t)E + E^{\tran}Q(t)A + C^{\tran}C -
                  E^{\tran}Q(t)BB^{\tran}Q(t)E\tag{ADRE}\label{eq:BKS20-ADRE}\\
  \intertext{\bfseries Non-autonomous differential Riccati equations}
  -{E(t)}^{\tran}\dot{Q}(t)E(t) &= {\left(A(t)+\dot{E}(t)\right)}^{\tran}Q(t)E(t)
         + {E(t)}^{\tran}Q(t)\left(A(t)+\dot{E}(t)\right)
           \tag{NDRE}\label{eq:BKS20-NDRE}\\ 
    &\phantom{=} + {C(t)}^{\tran}C(t) - 
                   {E(t)}^{\tran}Q(t)B(t){B(t)}^{\tran}Q(t)E(t)
\end{align*}
The last equations are formulated for the time-varying counterpart
of~\eqref{eq:BKS20-sys}, i.e.\ the system where all matrices are allowed to
depend on time as well. Both DREs contain the case of differential Lyapunov
equations. Optimized solvers for those are still work in progress and must at
the moment be implemented by setting either \(B\), or \(C\) (in the dual
equation) to zero and thus eliminating the quadratic term. Available solution
methods in \mmess{} are described in
Section~\ref{sec:BKS20-philosophy_features}.

Classic Lyapunov equation based balanced truncation is known to preserve
asymptotic stability of the original system in the reduced-order model. Other
balancing-based methods have been developed to preserve other properties like
passivity or contractivity. For these special balancing-type MOR methods, other
matrix equations need to be solved that do not have a tailored solver in
\mmess{}, yet.  Still, they can be reformulated into one of the types above. In
order to have a more complete picture what equations can be solved with the
current \mmess{}, we list them here, but get back to them in
Section~\ref{sec:BKS20-mor} and describe their reformulations into the special
cases above and why they can still be solved using \mmess{}.
\begin{align*}
  \intertext{\bfseries Positive real balancing}
  \begin{split}
    0 &= APE^{\tran} + EPA^{\tran} + 
       (EPC^{\tran}-B){(D+D^{\tran})}^{-1}{(EPC^{\tran}-B)}^{\tran}\\
    0 &= A^{\tran}QE + E^{\tran}QA +
    (E^{\tran}QB-C^{\tran}){(D+D^{\tran})}^{-1}{(E^{\tran}QB-C^{\tran})}^{\tran}
  \end{split}\tag{PRARE}\label{eq:BKS20-PRARE}\\
  \intertext{\bfseries Bounded real balancing}
  \begin{split}
    0 &= APE^{\tran} + EPA^{\tran} + BB^{\tran} +
    (EPC^{\tran} + BD^{\tran}) {(I-DD^{\tran})}^{-1}
    {(EPC^{\tran} + BD^{\tran})}^{\tran}\\
    0 &= A^{\tran}QE + E^{\tran}QA + C^{\tran}C+
    (E^{\tran}QB + C^{\tran}D) {(I-D^{\tran}D)}^{-1}
    {(E^{\tran}QB + C^{\tran}D)}^{\tran}
  \end{split}\tag{BRARE}\label{eq:BKS20-BRARE}\\
  \intertext{\bfseries Linear-quadratic Gaussian balancing}
  \begin{split}
    0 &= APE^{\tran} + EPA^{\tran} + BB^{\tran} -
    (EPC^{\tran} + BD^{\tran}) {(I+DD^{\tran})}^{-1}
    {(EPC^{\tran} + BD^{\tran})}^{\tran}\\ 
    0 &= A^{\tran}QE + E^{\tran}QA + C^{\tran}C -
    (E^{\tran}QB + C^{\tran}D) {(I+D^{\tran}D)}^{-1}
    {(E^{\tran}QB + C^{\tran}D)}^{\tran}
  \end{split}\tag{LQGARE}\label{eq:BKS20-LQGARE}\\
\end{align*}
\subsection{A brief history of \mmess{}}\label{sec:BKS20-history}
\paragraph{\bfseries Early days, the \lyapack{} years}
The package \mmess{} originates in the work of Penzl~\cite{BenLP08,Pen00b,Pen00}
around the year 2000. More precisely, we understand \mmess{} as a continuation
and successor of Penzl's \lyapack{}-toolbox~\cite{Pen00a} for \matlab{}. While
most of the basic ideas from the original package have been preserved, some
features have been abandoned and some have been altered to improve efficiency
and reliability.

The treatment of generalized state-space systems, i.e.\
systems~\eqref{eq:BKS20-sys} with nontrivial, i.e.\ non-identity, \(E\)-matrices
have been added first. These changes still happened under the \lyapack{}-label
in versions~1.1--1.8 until about 2007. 

\paragraph{\bfseries Transition to \mmess{} and present} 
The transition to the relabeled \mmess{}-1.0 package
included a complete reorganization of the process data. Also, \lyapack{} used
string manipulations and \texttt{eval}-calls to mimicked function pointers,
which 
we replaced by proper function handles supported in modern \matlab{} and
\octave{}. Moreover, the formulation of the low-rank alternating directions
implicit (LR-ADI) iteration, which always was the heart and soul of \lyapack{},
was greatly updated to allow for cheaper evaluation of stopping criteria and an
iteration inherent generation of real solution factors, which could only be
achieved through post-processing in \lyapack{}.

The necessity for an a priori selection of shift parameters for convergence
acceleration used to be a major point of criticism regarding the ADI based
solvers. The selection of shift generation methods was extended in \mmess{}
and especially a new method that automatically generates the shifts during the
iteration~\cite{Kue16} was added, which makes the solvers accessible also to
non-experts. 

Other than that, version 1.0 saw general code modernization to support optimized
features in \matlab{} and to be 100\% \octave{} compatible. 

The two major contributions of version 2.0 were the
inclusion of the RADI iteration\cite{BenBKetal18} for~\eqref{eq:BKS20-ARE} and
several solvers for 
differential Riccati equations in both the autonomous~\eqref{eq:BKS20-ADRE} and
non-autonomous~\eqref{eq:BKS20-NDRE} cases.

Moreover, over time more system classes, including specially structured
differential algebraic equation (DAE) based systems and second-order systems,
have been added.

\paragraph{\bfseries  Future development plans} 
The most immediate upcoming feature in the near future is the inclusion of
Krylov subspace projection methods for algebraic Lyapunov~\cite{Sim07, SimD09}
and Riccati equations~\cite{SimSM14,LinS15,Sim16}. The infrastructure and
solvers are under current development and the feature is going to be part of
version 3.0. The plans for the more distant future include, inclusion of
low-rank solvers for Sylvester equations~\cite{BenK14} and non-symmetric
AREs~\cite{BenKS16}, as well as the discrete-time counterparts of the existing
equations, i.e.\ Stein equations~\cite{JbiM11,LiWCetal13,BenK14,Kue16,PonK19}
and discrete time Riccati equations. Also, more complex sets of equations like
Lur'e equations~\cite{PolR12} and Lyapunov-plus-positive
equations~\cite{morBre13,morBenG17,ShaSS16} are currently investigated and will
be added if solvers can be implemented in a robust and efficient way using the
\mmess{} infrastructure.
\subsection{Structure of this chapter}\label{sec:BKS20-struct}
The following section introduces \mmess{} and its basic implementation
philosophy. It further elaborates on supported system structures beyond the
basic form in~\eqref{eq:BKS20-sys} and describes current basic features of the
package. Section~\ref{sec:BKS20-mor} is dedicated to the description of MOR
methods contained or demonstrated in \mmess{}, while
Section~\ref{sec:BKS20-pmor} shows how the existing tools in \mmess{} can be
used to implement basic PMOR methods from the literature. The last section
demonstrates how \mmess{} can be employed in parametric model order reduction,
solving a selection of the above equations, for the benchmark example introduced
in the separate chapter~\cite{morRavS20} of this volume. Similarly, this benchmark setting is
considered in the other software chapters~\cite{morBenW20c,morMliRS20,morHim20},
in order to compare the applicability of the individual packages in a
standardized setting.
\section{\mmess{} --- Philosophy and Features}%
\label{sec:BKS20-philosophy_features}
\begin{table}[tbp]
  \centering
     \begin{tabularx}{\linewidth}{|l|L|L|L|L|L|}
       \hline
       usfs &
       {\ttfamily default} &
       {\ttfamily so\_1 / so\_2}&
       {\ttfamily dae\_1}&
       {\ttfamily dae\_2}&
       {\ttfamily dae\_1/2/3\_so}
       \\\hline\hline
       System &
       standard / generalized state-space form&
       second-order 1st~/~2nd companion form&
       semi-explicit index-1 DAE&
       semi-explicit index-2 Stokes-type DAEs&
       semi-explicit second-order index-1/2/3 DAEs using companion form
       \\\hline
       Demos &
       FDM~\cite{Pen00a}, Rail~\cite{morBenS05} &
       TripleChain~\cite{TruV09,Saa09} &
       DAE1 (BIPS Power-systems model~\cite{morFreRM08}) &
       DAE2 Stokes~\cite{morSch07}, K\'arman vortex shedding~\cite{Wei16} &
       constrained TripleChain
       \\\hline
     \end{tabularx}
  \caption{Supported system structures via user-supplied function sets (usfs).}%
  \label{tab:BKS20-usfs}
\end{table}
\begin{table}[tbp]
  \renewcommand\tabularxcolumn[1]{m{#1}}
  \def\arraystretch{1.75}\setlength\tabcolsep{1ex}
  \centering
  \begin{tabularx}{\linewidth}{|>{\ttfamily}l|X|}
    \hline
    \textrm{function call}&operation\\
    \hline\hline
    Y = oper.mul\_A(eqn,opts,opA,B,opB) & 
        \(Y = A^{\texttt{opA}}B^{\texttt{opB}}\)\\\hline
    Y = opr.mul\_E(eqn,opts,opE,B,opB) & 
        \(Y = E^{\texttt{opE}}B^{\texttt{opB}}\)\\\hline
    Y = oper.mul\_ApE(eqn,opts,opA,p,opE,B,opB) & 
        \(Y = \left(A^{\texttt{opA}}+ p E^{\texttt{opE}}\right) B^{\texttt{opB}}\)
    \\\hline\hline
    X = oper.sol\_A(eqn,opts,opA,B,opB) & 
        \( A^{\texttt{opA}}X = B^{\texttt{opB}}\)\\\hline
    X = oper.sol\_E(eqn,opts,opE,B,opB) & 
        \( E^{\texttt{opE}}X = B^{\texttt{opB}}\)\\\hline
    X = oper.sol\_ApE(eqn,opts,opA,p,opE,B,opB) & 
        \(\left(A^{\texttt{opA}}+ p E^{\texttt{opE}}\right) X = B^{\texttt{opB}}\)
    \\\hline\hline
    result = oper.init(eqn,opt,oper,f1,f2) & 
      general initialization and sanity checks\\\hline
    [W,res0] = oper.init\_res(eqn,opts,oper,V) & 
    Compute initial residual factor \texttt{W} from \texttt{V}, and
                                                 \(\texttt{res0}=\nrm{W}\)
    \\\hline\hline
    \parbox[l][2.5em][b]{.6\textwidth}{[eqn,opts,oper] = \newline
    eval\_matrix\_functions(eqn,opts,oper,t)}& 
     In the time-varying case, fix all
    the above to time instance \texttt{t}.
    \\\hline\hline
    n = oper.size(eqn,opts,oper) & 
      Returns the dimension \(n\) in~\eqref{eq:BKS20-sys}.
    \\\hline
  \end{tabularx}
  \caption{User supplied function names and their actual operation.}%
  \label{tab:BKS20-usf-list}
\end{table}
The \mmess{} philosophy relies on three simple principles:
\begin{description}
\item[\bfseries abstract state-space system] All routines assume to work on a
  system of the form~\eqref{eq:BKS20-sys}, or~\eqref{eq:BKS20-sysinf}. For a
  simple spatially discretized parabolic PDE,~\eqref{eq:BKS20-sys} is exactly
  given by the sparse matrices describing the semi-discretized system. For other
  systems,~\eqref{eq:BKS20-sys} may be a dense, inaccessible realization, like,
  e.g.\ a projection to a hidden manifold for a Stokes-type DAE system.
\item[\bfseries implicit reformulation] When the system matrices are potentially
  dense or even inaccessible, or otherwise prohibitive to use, the matrices are
  never formed explicitly, but only their actions are expressed in terms of the
  original data. For the aforementioned DAEs this means, only the given
  semi-explicit system matrices are employed, but the algorithm runs as if it
  was formulated on the hidden manifold, i.e.\ for the implicit ordinary
  differential equation. This technique is often also called {\em implicit
  index-reduction}. For second-order systems, similarly, it is sometimes
  prohibitive to work with the double-sized phase-space realization in companion
  form. Again, all operations are executed only using the original
  second-order matrices, while solutions live in the double sized space.
\item[\bfseries operation abstraction] The abstraction of operations is realized
  via the so-called user-supplied function sets (usfs), which we have inherited
  from \lyapack{}. In comparison to \lyapack{} we have slightly extended
  this set of functions. At the same time, we have removed the necessity to
  provide empty functions, which are now automatically replaced by a
  \texttt{do\_nothing} function. While making things far more complicated in,
  e.g.\ the \texttt{default} case (see Table~\ref{tab:BKS20-usfs}), where all
  matrices are expected to be available, this allows to hide the actual matrix
  realization from the algorithms. This way, in principle, the algorithms can
  run matrix-free with respect to \(A\) and \(E\) as demonstrated
  in~\cite{BenSSetal14}.
\end{description}

The basic structure and design, of \mmess{}, was decided when object-oriented
features in \matlab{} were in their early stages and essentially absent in
\octave{}. Still some of the design follows object-oriented paradigms. We mimic
the object-orientation by passing three central data-structures through all
relevant functions. These three items of type \texttt{struct} are
\begin{description}
\item[\ttfamily eqn] This structure essentially holds all relevant information
  about the underlying system~\eqref{eq:BKS20-sys}, or~\eqref{eq:BKS20-sysinf}
  and determines which equation in the dual pair we are aiming to solve, by
  \texttt{eqn.type='N'}, or \texttt{eqn.type='T'} representing the transposition
  on the left multiplication by \(A\). 
\item[\ttfamily oper] The operator structure, generated by the function
  \texttt{operatormanager}, holds all function handles for the relevant
  operations with the system matrices \(A\) and \(E\). A list of these
  operations can be found in Table~\ref{tab:BKS20-usf-list}. Most functions in
  the list are accompanied by two functions, with appendices \texttt{\_pre} and
  \texttt{\_post}, called at the beginning and the end of a function working with
  them. They are intended for the generation and clean up of helper data, like
  the pre-factorization of matrices, when a sparse direct solver is used, or the
  generation of a preconditioner for an iterative solver.
\item[\ttfamily opts] The actual options structure is a structure of structures,
  i.e.\ it has a substructure for each algorithm/function, but also holds
  central information on the top level. For example \texttt{opts.norm} defines
  the norm that should consistently be used in all operations and hierarchy
  levels of the potentially cascaded algorithms, while
  substructures like \texttt{opts.adi}, or \texttt{opts.shifts} provide the
  specific control parameters for the LR-ADI algorithm and the shift
  computation.   
\end{description}
Note that for all matrix operations in the usfs, we allow for corresponding
\texttt{\_pre} and \texttt{\_post} functions. Other functions like \texttt{init}
or \texttt{size} do not support \texttt{\_pre} and \texttt{\_post}.

While the function handles in \texttt{oper} work on the original \(A\)
from~\eqref{eq:BKS20-sys}, sometimes it is necessary to actually work with
low-rank updated versions of \(A\) in the form \(A + UV^{\tran}\). We have seen
an example in~\eqref{eq:BKS20-iARE}, where \(\tA\) is in the very
form. Another prominent appearance is the Newton-Kleinman iteration
(see~\cite{Kle68} for classic iteration and~\cite{BenLP08} for the low-rank
version) for~\eqref{eq:BKS20-ARE}, where in iteration \(j\), the step equation
(for the second equation in the pair) takes the form
\[
  {\left(A -BK_{j-1}\right)}^{\tran}X_{j} E + E^{\tran}  X_{j} \left(A
    -BK_{j-1}\right) = {\left[C\; K_{j-1}\right]}^{\tran}\left[C\; K_{j-1}\right].
\]
Therefore, most solvers in \mmess{} assume that this structure can be given. The
flag \texttt{eqn.haveUV} set to a non-zero value indicates that this is the
case. Then the fields \texttt{eqn.U} and \texttt{eqn.V} need to hold the
corresponding dense rectangular matrices of compatible dimensions. Similarly, the
field \texttt{eqn.haveE} indicates that a non-trivial, i.e.\ non-identity \(E\)
matrix is present and needs to be used via the function handles in
Table~\ref{tab:BKS20-usf-list}.

Note that it is prohibitive to form \(A + UV^{\tran}\) explicitly, since even
for very sparse \(A\) it can easily be a dense matrix. Especially, it is
prohibitive to use direct solvers based on matrix decompositions on it,
since then even if \(A + UV^{\tran}\) manages to preserve some sparsity, the
fill-in will make the triangular factors dense. Therefore, all linear systems
with \(A + UV^{\tran}\) are solved via the Sherman-Morrison-Woodbury
matrix-inversion formula (see, e.g.\ \cite[Section 2.1.4]{GolV13}) in \mmess{}.
\subsection{Available solver functions and underlying methods}%
\label{sec:BKS20-avail-solv-funct}
We provide two solvers for the standard cases in~\eqref{eq:BKS20-ALE}
and~\eqref{eq:BKS20-ARE} that are purely matrix-based, intended for large-scale
sparse matrix coefficients and classic 2-term low-rank factorizations of the
constant terms and cores in the quadratic terms. The functions are called
\texttt{mess\_lyap} and \texttt{mess\_care} and mimic the calls of \texttt{lyap}
and \texttt{care} from \matlab{}'s control systems toolbox, or the \octave{}
control package, for dense matrices.

Other than that, we have the functions in Table~\ref{tab:BKS20-slovers} that
allow for more flexible tuning, solve a large variety of equations and,
especially, benefit from the full potential of the user supplied functions.  In
the table we give references to the most state of the art presentations of the
algorithms in the literature, on which our implementations are based.
\begin{table}[tbp]
  \renewcommand\tabularxcolumn[1]{m{#1}}
  \def\arraystretch{1.75}\setlength\tabcolsep{1ex}
  \begin{tabularx}{1.0\linewidth}{|>{\ttfamily}l|L|c|}
    \hline
    \textrm{solver}&description&reference\\
    \hline
    \multicolumn{3}{c}{\bfseries algebraic Lyapunov equations}\\
    \hline                                 
    mess\_lradi& The low-rank alternating directions implicit (LR-ADI) iteration
                 in residual based formulation and with automatic shift
                 selection for~\eqref{eq:BKS20-ALE}.
                               &\cite{Kue16}\\\hline
    \multicolumn{3}{c}{\bfseries algebraic Riccati equations}\\
    \hline 
    mess\_lrnm&An inexact Kleinman-Newton iteration with line search
         for~\eqref{eq:BKS20-ARE}.&\cite{Wei16}\\\hline 
    mess\_lrradi&The RADI iteration for~\eqref{eq:BKS20-ARE}
                &\cite{BenBKetal18} \\\hline
    mess\_lrri&A low-rank version of the Riccati iteration~\cite{LanFA07}
                for~\eqref{eq:BKS20-iARE}&\\\hline
    \multicolumn{3}{c}{\bfseries differential Riccati equations}\\
    \hline
    mess\_bdf\_dre&Low-rank formulation of backward differentiation formulas for
          large-scale differential Riccati
          equations~\eqref{eq:BKS20-ADRE},~\eqref{eq:BKS20-NDRE} 
                  &\cite{Lan17}\\\hline
    mess\_rosenbrock\_dre&Low-rank formulation of Rosenbrock
          methods for large-scale differential Riccati
          equations~\eqref{eq:BKS20-ADRE}
         &\cite{Lan17}\\\hline
    mess\_splitting\_dre&Splitting schemes for large-scale differential Riccati
          equations~\eqref{eq:BKS20-ADRE},~\eqref{eq:BKS20-NDRE}
         &\cite{Sti15,Sti18a}\\\hline
  \end{tabularx}
  \caption{Solver functions with algorithm and feature descriptions and latest
    and most feature complete literature references.}\label{tab:BKS20-slovers}
\end{table}

\section{Model Order Reduction in \mmess{}}\label{sec:BKS20-mor}
The basic model order reduction (MOR) facilities in \mmess{} are limited. Still,
all building blocks for projection-based MOR using balancing methods, where
matrix equations are most obviously applied, are available. For the sake of
completeness and to fix our notation, we repeat the basics of projection based
MOR.\ Given a state-space system of the form~\eqref{eq:BKS20-sys}, we search for
the two rectangular transformation matrices \(V,\,W\in\Rnr\) that define the
actual oblique projection \(T = V {(W^{\tran}V)}^{-1} W^{\tran}\), but transform
the system into the reduced coordinates directly. The reduced-order model then
takes the form
\begin{align*}
  \begin{split}
    \hE \dot\hx(t) &= \hA\hx(t) + \hB u(t),\\
    \hy(t) &= \hC\hx(t) + Du(t),
  \end{split}\label{eq:BKS20-ROM}\tag{ROM}
\end{align*}
where \(\hE = W^{\tran}EV,\,\hA = W^{\tran}AV\in\Rrr\),
\(\hB=W^{\tran}B\in\Rrm\), and \(\hC = CV\in\Rpr\).

The number of actual MOR routines in \mmess{} is rather limited. In version
2.0.1, we have \texttt{mess\_balanced\_truncation} implementing classic Lyapunov
balancing~\cite{morMoo81,morTomP87,morLauHPetal87}, for
systems~\eqref{eq:BKS20-sys} realized with sparse \(E\) and \(A\)
(\cite{BenLP08,morGugL05,Saa09}), and \texttt{mess\_tangential\_irka}
implementing the tangential iterative Krylov algorithm (IRKA)~\cite{morGugAB08}
for first and second order systems. Our Gramian computation methods are
integrated in a range of MOR software packages, though. While
\textrm{sssMOR}~\cite{morCasCJetal17} directly calls \mmess{}-1.0.1, for other
packages like \textrm{MOREMBS}~\cite{FehGHetal18} and
\textrm{MORPACK}~\cite{morpackweb} we have contributed tailored versions of our
algorithms.

Also, we provide tools like a square root method (SRM) function to compute the
transformation matrices \(V\) and \(W\) from given Gramian factors. This
function currently only uses the classic Lyapunov balancing error bound in the
adaptive mode. This is subject to change in future versions.
\subsection{IRKA and classic balanced truncation}%
\label{sec:BKS20-class-lyap-balanc}
Consider that all matrices in~\eqref{eq:BKS20-sys} are available. As an example
we use the Steel Profile benchmark~\cite{morBenS05,morwiki_steel}, included in
\mmess{}, using the version with \(n=1\,357\). Then computing the reduced order
matrices \(E_{r}\), \(A_{r}\), \(B_{r}\), \(C_{r}\) for maximum reduced order
\(25\) using the tangential IRKA~\cite{morGugAB08} is as easy as calling:
\begin{lstlisting}[language=Matlab]
eqn = getrail(1);
opts.irka.r = 25;
[Er, Ar, Br, Cr] = mess_tangential_irka(eqn.E_,eqn.A_, eqn.B, eqn.C, opts)
\end{lstlisting}
This will use default values for maximum iteration numbers and stopping
criteria, which can be refined via the \texttt{opts.irka} structure. For a list
of available options see \texttt{help mess\_tangential\_irka}.

Analogously, to compute a (Lyapunov) balanced truncation 
approximation of maximum order \(50\) and with an absolute
\(H_{\infty}\)-error tolerance of \(10^{-2}\) for the same model the simplest
call is: 
\begin{lstlisting}[language=Matlab]
eqn = getrail(1);
[Er, Ar, Br, Cr] = mess_balanced_truncation(eqn.E_, eqn.A_, eqn.B, eqn.C, 50, 1e-2);
\end{lstlisting}
Note that Lyapunov balancing leaves the \(D\) matrix untouched in general, while
it is absent in this example anyway. Note, further, that the interface may
change slightly in future releases to make it more consistent with that of the
IRKA function and to allow for the addition of the other balancing methods.

The balanced truncation approximation can be achieved in a step by step
procedure first computing the two Gramian factors, then applying them in the
square root method to determine \(V\) and \(W\), and finally compressing the
large-scale matrices to the reduced-order system matrices. This can all be
executed using the procedural building blocks of
\texttt{mess\_balanced\_truncation}. The example \texttt{bt\_mor\_rail\_tol} in
the \texttt{DEMOS/Rail} folder, residing in the main installation folder of
\mmess{}-2.0.1, demonstrates this procedure. The step-wise approach can also be
used for a number of structured systems like second-order systems, or
semi-explicit DAE systems, while \texttt{mess\_balanced\_truncation} only
supports generalized systems with invertible \(E\), and all coefficients given
explicitly as matrices, at the moment. See Table~\ref{tab:BKS20-btdemos} for an
overview of demonstration examples explaining these procedures.
\begin{table}[tbp]
  \renewcommand\tabularxcolumn[1]{m{#1}}
  \def\arraystretch{1.75}\setlength\tabcolsep{1ex}
  \begin{tabularx}{1.0\linewidth}{|>{\ttfamily}l|L|c|}
    \hline
    \textrm{Example} & Description & References\\
    \hline\hline
    bt\_mor\_DAE1\_tol & Balanced truncation for a semi-explicit power systems
                         model 
                      of differential index 1&\cite{morFreRM08}\\\hline
    bt\_mor\_DAE2 & Balanced truncation for Stokes and Oseen equations of index
                    2 
                                   &\cite{morHeiSS08, morBenSU16}\\\hline
    BT\_TripleChain & First-order and structure-preserving balanced truncation
                     for a model with three coupled mass-spring-damper chains
                   &\cite{TruV09,Saa09,morReiS08}\\\hline
    BT\_sym\_TripleChain & As above, but exploiting state-space symmetry of the
                         tailored companion form first-order reformulation
                       &\\\hline
    BT\_DAE3\_SO & First-order and structure-preserving balanced truncation for
                   a 
                 variant of the above system that has a constraint turning it 
                 into an index-3 DAE &\cite{morSaaV18,morUdd15,Udd19}\\\hline
  \end{tabularx}
  \caption{Demonstration examples for balanced truncation of structured systems
    in \mmess{}.}\label{tab:BKS20-btdemos}
\end{table}

\subsection{Further variants of balanced truncation}%
\label{sec:BKS20-var-balanc}
We have shown the Riccati equations defining the Gramians employed in
positive-real, bounded-real, and linear-quadratic Gaussian balanced truncation
in equations~\eqref{eq:BKS20-PRARE}, \eqref{eq:BKS20-BRARE},
\eqref{eq:BKS20-LQGARE} 
in the Introduction. Assuming, we have computed the Gramian factors, the
reduced-order models can be derived, along the lines of the demonstration
examples from Table~\ref{tab:BKS20-btdemos}. This can be done using the same
\mmess{} function at least for a fixed desired reduced order. The error bound
based 
order decision in the square root method needs adaptation to the specific error
bound in some cases, though, see e.g.\ \cite[Section~7.5]{morAnt05} for a
comparison of the bounds and procedures.

Here, we restrict ourselves to presenting how the specially structured Riccati
equations can be solved with the existing functionality in \mmess{}.
\subsubsection{Positive-real balancing}\label{sec:BKS20-prbt}
For positive-real systems, by definition \(D+D^{\tran}\) is positive-definite,
when it is invertible.  This is always the case when the Riccati equations exist
and do not degenerate to a set of Lur'e equations. Then we can decompose
\(D+D^{\tran}\) into Cholesky factors, i.e.\ \(R^{\tran}R=D+D^{\tran}\). Using
these Cholesky factors, we define
\[
  \tE = E, \qquad
  \tA= A + UV^{\tran},\qquad
  \tB = B R^{-1},\qquad
  \tC = R^{-\tran} C,
\]
with \(U=\tB\) and \(V^{\tran}= \tC\),
and a straight forward calculation shows that~\eqref{eq:BKS20-PRARE} can be
rewritten in the form 
\begin{align*}
  0 &= \tA P\tE^{\tran} + \tE P\tA^{\tran} + \tB\tB^{\tran} + \tE P\tC^{\tran}\tC
      P\tE^{\tran}\\ 
  0 &= \tA^{\tran}Q\tE + \tE^{\tran}Q\tA  + \tC^{\tran}\tC +
      \tE^{\tran}Q\tB\tB^{\tran}Q\tE.
\end{align*}
This resembles the Riccati case in~\eqref{eq:BKS20-iARE} with a low-rank updated
matrix \(A\) and only the positive quadratic term present. This case is
supported by the \texttt{mess\_lrri} routine. Note that \(D+D^{\tran}\) is of
small dimension, such that this reformulation is always feasible.

\subsubsection{Bounded-real balancing}\label{sec:BKS20-brbt}
The bounded-real assumptions guarantee that \(I-DD^{\tran}\) and
\(I-D^{\tran}D\) are symmetric positive  definite. Therefore, we can decompose
them into Cholesky factors, i.e.\ \(R^{\tran}R = I - DD^{\tran}\) and
\(L^{\tran}L = I - D^{\tran}D\). Now, we define
\[
  \tE = E, \qquad
  \tA= A + UV^{\tran},\qquad
  \tB = B L^{-1},\qquad
  \tC = R^{-1} C,
\]
with \(U=B D^{\tran}\) and \(V^{\tran}=  {(I - DD^{\tran})}^{-1} C\)
and another technical, but straight forward, calculation shows
that~\eqref{eq:BKS20-BRARE} can be rewritten in the form 
\begin{align*}
  0 &= \tA P\tE^{\tran} + \tE P\tA^{\tran} + \tB\tB^{\tran}
      + \tE P\tC^{\tran}\tC P\tE^{\tran}\\ 
  0 &= \tA^{\tran}Q\tE + \tE^{\tran}Q\tA  + \tC^{\tran}\tC +
      \tE^{\tran}Q\tB\tB^{\tran}Q\tE.
\end{align*}
This, again, falls into the class of equations in~\eqref{eq:BKS20-iARE} with a
low-rank updated matrix \(A\) and only the positive square term present. As
mentioned above, this case is supported by the \texttt{mess\_lrri} routine. For
the same reason as above, this reformulation can always be done.
\subsubsection{Linear-quadratic Gaussian balancing}\label{sec:BKS20-lqgbt}
For linear-quadratic Gaussian balanced truncation, an important special case
(see, e.g.\ \cite{morBenH15,morMusG91,morMoeRS11,BatHEetal06}) is
\(D=0\). In that
case~\eqref{eq:BKS20-LQGARE} obviously reduces to the standard Riccati
equation~\eqref{eq:BKS20-ARE} that can be solved using \texttt{mess\_lrnm} or
\texttt{mess\_lrradi}. The corresponding \mmess{} workflow is demonstrated in
the \texttt{lqgbt\_mor\_FDM} example for a simple heat equation model
semi-discretized by the finite difference method.

On the other hand, when \(D\not=0\), it is, by standard assumptions in \mmess{},
real and all eigenvalues of \(DD^{\tran}\) and \(D^{\tran}D\) are
non-negative. Therefore, \(I+DD^{\tran}\) and \(I+D^{\tran}D\) are symmetric and
positive definite and analogous to the above, we can decompose into Cholesky
factorizations \(R^{\tran}R = I + DD^{\tran}\) and
\(L^{\tran}L = I + D^{\tran}D\). We now define
\[
  \tE = E, \qquad
  \tA= A + UV^{\tran},\qquad
  \tB = B L^{-1},\qquad
  \tC = R^{-1} C,
\]
with \(U=- B D^{\tran}\) and \(V^{\tran}={(I+DD^{\tran})}^{-1} C\).
An analogous calculation to the bounded-real case shows
that~\eqref{eq:BKS20-LQGARE} can be rewritten in the form
\begin{align*}
  0 &= \tA P\tE^{\tran} + \tE P\tA^{\tran} + \tB\tB^{\tran} 
      - \tE P\tC^{\tran}\tC P\tE^{\tran}\\ 
  0 &= \tA^{\tran}Q\tE + \tE^{\tran}Q\tA  + \tC^{\tran}\tC 
      - \tE^{\tran}Q\tB\tB^{\tran}Q\tE.
\end{align*}
Due to the differing signs, here, we end up with a standard Riccati
equation~\eqref{eq:BKS20-ARE}, just like in the case \(D=0\). Again the
transformation is always feasible in the sense of \mmess{} applicability.
\section{Parametric Model Order Reduction using \mmess{}}%
\label{sec:BKS20-pmor} 
Parametric MOR (PMOR) aims to preserve symbolic parameters in the original system
description also in the reduced-order model. In the most general case that means
the system 
\begin{equation*}
  \begin{split}
    E(\mu)\dot x(\mu,t) &= A(\mu) x(\mu,t) + B(\mu) u(t),\\
    y(\mu,t) &= C(\mu) x(\mu,t) + D(\mu) u(t),
  \end{split}\label{eq:BKS20-musys}\tag{\(\Sigma(\mu)\)}
\end{equation*}
is transformed into 
\begin{align*}
  \begin{split}
    \hE(\mu) \dot\hx(\mu,t) &= \hA(\mu) \hx(\mu,t) +
    \hB(\mu) u(t),\\
    \hy(t) &= \hC(\mu)\hx(\mu,t) + D(\mu)u(t).
  \end{split}\label{eq:BKS20-muROM}\tag{ROM(\(\mu\))}
\end{align*}

By default, \mmess{}-2.0.1 does not support PMOR.\  It is, however, very easy to
implement basic PMOR routines building up on the methods from the previous
section. The key ingredient, that at the same time establishes the link to the
previous section in many methods, is the necessity to evaluate standard MOR
problems in certain training points for given parameter values \(\mu^{(i)}\)
(\(i=1,\dots,k\)), e.g.\ on a sparse-grid in the parameter domain.  
While piecewise MOR approaches (e.g.\ \cite{morBauBBetal11}) 
aim to find constant global (with respect to the parameter)
transformation matrices \(V\) and \(W\) to derive~\eqref{eq:BKS20-muROM}, other
methods aim to establish it by interpolation of some kind. The literature
basically 
provides three approaches interpolating different system features, see,
e.g.,~\cite{morBenGW15} for further categorization of PMOR methods: 
\begin{itemize}
\item matrix interpolation, i.e.\ function
  interpolation of the parameter-dependent coefficient matrices, or the
  transformation matrices, (e.g.\ \cite{morPanMEetal10, morGeuBPetal14,
    morGeuPWetal12, morAmsF11}),  
\item interpolation of the transfer functions in the parameter
  variable~\cite{morBauB09},
\item interpolation of system poles (e.g.\ \cite{morBenGH15,morYueFB19a}). 
\end{itemize}

We demonstrate the basic steps for piecewise and interpolatory methods along the
lines of~\cite{morBauBBetal11,morBauB09} in the remainder of this section and
give numerical illustrations in Section~\ref{sec:BKS20-Numerics}.
\subsection{Piecewise MOR }\label{sec:BKS20-piecewise-mor}
We have mentioned above that the aim, here, is to find \(V\) and \(W\) constant,
such that \(\hE(\mu)=W^{\tran}E(\mu)V\), \(\hA(\mu)=W^{\tran}A(\mu)V\),
\(\hB(\mu)=W^{\tran}B(\mu)\), \(\hC(\mu)=C(\mu)V\). The strong point of this
method is that it trivially allows the ROMs in the parameters \(\mu^{(i)}\)
to vary in their reduced-order. This is due to the fact that
\begin{equation*}
  V = \left[V^{(1)}\cdots V^{(k)}\right]\qquad\text{and}\qquad
  W = \left[W^{(1)}\cdots W^{(k)}\right],
\end{equation*}
with \(V^{(i)}\) and \(W^{(i)}\) the transformation matrices at parameter sample
\(\mu^{(i)}\). 
This concatenation should be followed by a rank truncation to eliminate linear
dependencies. 

It essentially does not matter how the single transformation matrices have been
generated. We follow the presentation in~\cite{morBauBBetal11}, where IRKA is
used. In the numerical experiments we also compare to versions using balanced
truncation in the training samples.
\subsection{Interpolation of transfer functions}%
\label{sec:BKS20-interp-mor}
The representation of~\eqref{eq:BKS20-musys} in frequency domain after Laplace
transformation in the time variable (\(t\)), leads to the transfer function 
\begin{equation*}
  H(\mu,s)  = C(\mu){\left(sE(\mu) -A(\mu)\right)}^{-1}B(\mu).
\end{equation*}
For fixed \(\mu\), the IRKA method seeks to interpolate this function in
\(s\)-direction, while the well known balanced truncation method computes an
approximation to this function, with a computable error bound.
Therefore, it is an obvious task to try
to extend these features by interpolation in \(\mu\)-direction. Baur and
Benner meet this goal in~\cite{morBauB09} for local balanced truncation
approximations of~\eqref{eq:BKS20-musys}, achieving both stability preservation
and an error bound, i.e.\ the selling points of balanced truncation. Moreover
their method shares the flexibility with respect to the ROM orders, since the
interpolation is done via the transfer function, that has fixed dimension
independent of the realization of the system. On the other hand, interpolation
on matrix manifolds and with respect to system invariants need to fix the
dimensions of those objects.

For simplicity we restrict ourselves to the case of scalar parameters. The
approach in~\cite{morBauB09} defines~\eqref{eq:BKS20-muROM} via its transfer
function, which is chosen as an interpolant of the form
\begin{align*}
  \hH(\mu,s) = \sum_{i=1}^{k}\ell_{i}(\mu)\hH^{(i)}(s)
  &=\sum_{i=1}^{k} \ell_{i}(\mu)
    \hC^{(i)}{\left(s\hE^{(i)}-\hA^{(i)}\right)}^{-1}\hB^{(i)}\\ 
  &=\sum_{i=1}^{k} \hC^{(i)}(\mu)
    {\left(s\hE^{(i)}-\hA^{(i)}\right)}^{-1}\hB^{(i)}
\end{align*}
with scalar coefficients functions \(\ell_{i}(\mu)\), \(\hH^{(i)}(s)\) the
transfer function of the ROM at parameter sample \(\mu^{(i)}\) and
\(\hC^{(i)}(\mu) = \ell_{i}(\mu) \hC^{(i)}\). One can use the
last identity to define the matrices for the ROM-realization
\begin{align*}
  \hE &= 
  \begin{bmatrix}
    \hE^{(1)}&&\\
    &\myddots&\\
    &&\hE^{(k)}
  \end{bmatrix},
  \qquad
  \hA = 
  \begin{bmatrix}
    \hA^{(1)}&&\\
    &\myddots&\\
    &&\hA^{(k)}
  \end{bmatrix},
  \qquad
  \hB = 
  \begin{bmatrix}
    \hB^{(1)}\\
    \myvdots\\
    \hB^{(k)}
  \end{bmatrix},\\
  \hC(\mu) &= 
  \begin{bmatrix}
    \hC^{(1)}(\mu) & \cdots & \hC^{(k)}(\mu)
  \end{bmatrix},  
\end{align*}
such that
\begin{equation*}
  \hH(\mu,s) = \hC(\mu) {\left(s\hE -\hA\right)}^{-1}\hB.
\end{equation*}

Note that the parameter could as well be put into \(\hB\).
The specific choice of Lagrange polynomials is not necessary. We present
experiments with both classic polynomial interpolation and spline interpolation
in the next section.  Since we are dealing with scalar coefficient functions
here, it is advisable for a modern \matlab{}-implementation to exploit the power
of Chebfun~\cite{DriHT14,ChebfunWeb}. We do this for the polynomial
interpolation and the generation of the grid of training parameters, while the
splines use our own implementation.
 
\section{Numerical Experiments}\label{sec:BKS20-Numerics}
\begin{figure}[tbp]
  \centering
  \begin{subfigure}[t]{.49\textwidth} 
    \centering
    \definecolor{Gamma_in}{HTML}{ca0020}
\definecolor{Gamma_N}{HTML}{92c5de}
\definecolor{Gamma_D}{HTML}{0571b0}

\begin{tikzpicture}[x=.5\linewidth, y=.5\linewidth,thick]
  \foreach \x in {0.2,0.4,0.6,0.8} 
  {
    \draw[loosely dotted] (\x,0)--(\x,1);
    \draw[loosely dotted] (0,\x)--(1,\x);
    \draw (\x,0)--(\x,-.025);
    \draw (0,\x)--(-.025,\x);
    \node[below] at (\x, -.05) {\x};
    \node[left] at (-.05,\x) {\x};
  }
  \node[below left=.025] at (0,0) {(0,0)};
  \node[above left=.025] at (0,1) {(0,1)};
  \node[below right=.025] at (1,0) {(1,0)};
  \node[above right=.025] at (1,1) {(1,1)};
  \node at (.5,.5){\large\(\Omega_{0}\)};
  \foreach \x/\y/\n in { .3/.3/1, .7/.3/2, .7/.7/3, .3/.7/4 }
     {
     \draw[black] (\x,\y) circle [radius=.1];
     \node at (\x,\y) {\(\Omega_{\n}\)};
     }
  \foreach \x/\y/\n in { .1/.5/in, .5/.075/N, .5/1.05/N, 1.1/.5/D}
    \node[color=Gamma_\n] at (\x,\y) {\large\(\Gamma_{\n}\)};
  
  \draw[Gamma_in] (0,0)--(0,1);
  \draw[Gamma_N] (0,0)--(1,0);
  \draw[Gamma_N] (0,1)--(1,1);
  \draw[Gamma_D] (1,0)--(1,1);
  \foreach \x/\y/\n in { 0/0, 1/0, 0/1, 1/1}
    \filldraw[black] (\x,\y) circle [radius=1pt];

\end{tikzpicture}
    \caption{Computational domain and boundaries.}%
    \label{fig:BKS20-FOM-domain}
  \end{subfigure}
  \hfill
  \begin{subfigure}[t]{.49\textwidth} 
    \centering
  \tikzexternalenable%
  \tikzsetnextfilename{FOM_sigma}%
  \filemodCmp{graphics/FOM_sigma.tikz}{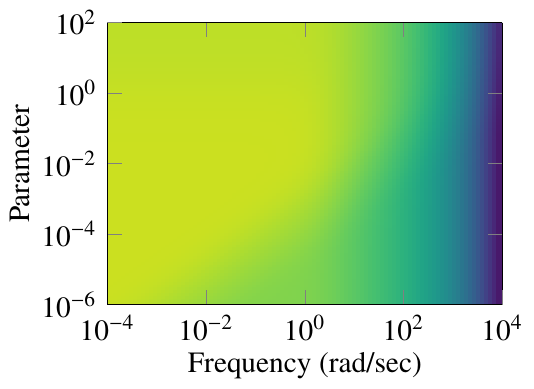}%
  {\tikzset{external/remake next}}{}%
  \begin{tikzpicture} 
  \pgfplotstableread{graphics/data/FOM_fom.dat}\tableROM 

  \begin{loglogaxis}[ 
    view={0}{90}, 
    width = .7\textwidth, 
    height = .5\textwidth, 
    scale only axis, 
    xmin = 1e-4, 
    xmax = 1e+4, 
    ymin = 1e-6, 
    ymax = 1e+2, 
    xtick = {1e-4, 1e-2, 1e0, 1e+2, 1e+4}, 
    ytick = {1e-6, 1e-4, 1e-2, 1e+0, 1e+2}, 
    zmode = log, 
    log base z = 10, 
    point meta min = -10, 
    point meta max = 1, 
    mesh/ordering = y varies, 
    mesh/rows = 100, 
    mesh/cols = 100, 
    xlabel = {\small Frequency (rad/sec)}, 
    xlabel style = {yshift = .3em}, 
    ylabel = {\small Parameter}, 
    ylabel style = {yshift = -.3em}, 
    scaled x ticks = false, 
    x tick label style = {/pgf/number format/fixed}] 

    \addplot3[surf, shader = flat] 
    table[x index = 0, y index = 1, z index = 2] {\tableROM}; 

  \end{loglogaxis} 
\end{tikzpicture} %
  \tikzexternaldisable%

    \caption{Sigma magnitude plot.}%
    \label{fig:BKS20-FOM-sigma}
  \end{subfigure}
  \vspace{.5\baselineskip} 

  \tikzexternalenable%
  \tikzsetnextfilename{FOM_cbar}%
  \filemodCmp{graphics/FOM_cbar.tikz}{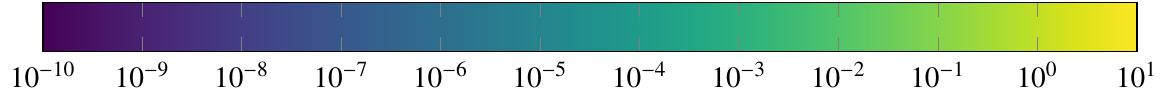}%
  {\tikzset{external/remake next}}{}%
  \begin{tikzpicture} 
  \begin{axis}[%
    hide axis, 
    scale only axis, 
    width = .95\textwidth,
    height = 1em, 
    point meta min = -10, 
    point meta max = 1, 
    colorbar, 
    colorbar horizontal, 
    colorbar style = { 
      xticklabel = $10^{\pgfmathparse{\tick} 
        \pgfmathprintnumber\pgfmathresult}$, 
      at = {(.5, 0)}, 
      anchor = north}, 
    scaled x ticks = false, 
    x tick label style = {/pgf/number format/fixed}] 
  \end{axis} 
\end{tikzpicture} %
  \tikzexternaldisable%

  \caption{Computational domain and sigma magnitude plot for the thermal block
    model.}\label{fig:BKS20-FOM}
\end{figure}
\begin{figure}[tbp] 
  \centering
  \begin{subfigure}[t]{.49\textwidth} 
    \centering 
  \tikzexternalenable%
  \tikzsetnextfilename{piecewise_bt_tol_err_rel}%
  \filemodCmp{graphics/piecewise_bt_tol_err_rel.tikz}{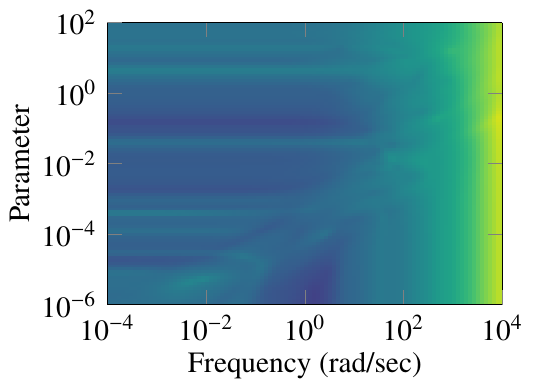}%
  {\tikzset{external/remake next}}{}%
  \begin{tikzpicture} 
  \pgfplotstableread{graphics/data/piecewise_BT_tol_rom_all.dat}\tableROM 

  \begin{loglogaxis}[ 
    view={0}{90}, 
    width = .7\textwidth, 
    height = .5\textwidth, 
    scale only axis, 
    xmin = 1e-4, 
    xmax = 1e+4, 
    ymin = 1e-6, 
    ymax = 1e+2, 
    xtick = {1e-4, 1e-2, 1e0, 1e+2, 1e+4}, 
    ytick = {1e-6, 1e-4, 1e-2, 1e+0, 1e+2}, 
    zmode = log, 
    log base z = 10, 
    point meta min = -12, 
    point meta max = 4, 
    mesh/ordering = y varies, 
    mesh/rows = 100, 
    mesh/cols = 100, 
    xlabel = {\small Frequency (rad/sec)}, 
    xlabel style = {yshift = .3em}, 
    ylabel = {\small Parameter}, 
    ylabel style = {yshift = -.3em}, 
    scaled x ticks = false, 
    x tick label style = {/pgf/number format/fixed}] 

    \addplot3[surf, shader = flat] 
    table[x index = 0, y index = 1, z index = 4] {\tableROM}; 

  \end{loglogaxis} 
\end{tikzpicture} %
  \tikzexternaldisable%
 
    \subcaption{Piecewise BT(\(10^{-4}\)) approximation.}
  \end{subfigure}%
  \hfill%
  \begin{subfigure}[t]{.49\textwidth} 
    \centering
  \tikzexternalenable%
  \tikzsetnextfilename{piecewise_tbt_tol_err_rel}%
  \filemodCmp{graphics/piecewise_tbt_tol_err_rel.tikz}{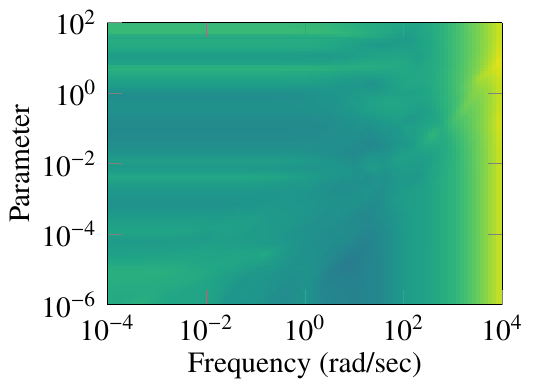}%
  {\tikzset{external/remake next}}{}%
  \begin{tikzpicture} 
  \pgfplotstableread{graphics/data/piecewise_BT_tol_trom_all.dat}\tableROM 

  \begin{loglogaxis}[ 
    view={0}{90}, 
    width = .7\textwidth, 
    height = .5\textwidth, 
    scale only axis, 
    xmin = 1e-4, 
    xmax = 1e+4, 
    ymin = 1e-6, 
    ymax = 1e+2, 
    xtick = {1e-4, 1e-2, 1e0, 1e+2, 1e+4}, 
    ytick = {1e-6, 1e-4, 1e-2, 1e+0, 1e+2}, 
    zmode = log, 
    log base z = 10, 
    point meta min = -12, 
    point meta max = 4, 
    mesh/ordering = y varies, 
    mesh/rows = 100, 
    mesh/cols = 100, 
    xlabel = {\small Frequency (rad/sec)}, 
    xlabel style = {yshift = .3em}, 
    ylabel = {\small Parameter}, 
    ylabel style = {yshift = -.3em}, 
    scaled x ticks = false, 
    x tick label style = {/pgf/number format/fixed}] 

    \addplot3[surf, shader = flat] 
    table[x index = 0, y index = 1, z index = 4] {\tableROM}; 

  \end{loglogaxis} 
\end{tikzpicture} %
  \tikzexternaldisable%
 
    \subcaption{Truncated piecewise BT(\(10^{-4}\)) approximation.}
  \end{subfigure} 

  \begin{subfigure}[t]{.49\textwidth} 
    \centering 
  \tikzexternalenable%
  \tikzsetnextfilename{piecewise_bt_fixed_err_rel}%
  \filemodCmp{graphics/piecewise_bt_fixed_err_rel.tikz}{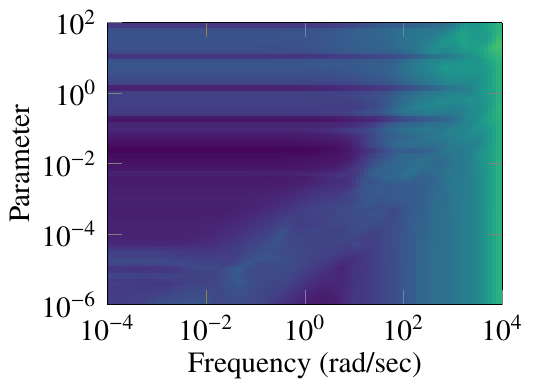}%
  {\tikzset{external/remake next}}{}%
  \begin{tikzpicture} 
  \pgfplotstableread{graphics/data/piecewise_BT_fixed_rom_all.dat}\tableROM 

  \begin{loglogaxis}[ 
    view={0}{90}, 
    width = .7\textwidth, 
    height = .5\textwidth, 
    scale only axis, 
    xmin = 1e-4, 
    xmax = 1e+4, 
    ymin = 1e-6, 
    ymax = 1e+2, 
    xtick = {1e-4, 1e-2, 1e0, 1e+2, 1e+4}, 
    ytick = {1e-6, 1e-4, 1e-2, 1e+0, 1e+2}, 
    zmode = log, 
    log base z = 10, 
    point meta min = -12, 
    point meta max = 4, 
    mesh/ordering = y varies, 
    mesh/rows = 100, 
    mesh/cols = 100, 
    xlabel = {\small Frequency (rad/sec)}, 
    xlabel style = {yshift = .3em}, 
    ylabel = {\small Parameter}, 
    ylabel style = {yshift = -.3em}, 
    scaled x ticks = false, 
    x tick label style = {/pgf/number format/fixed}] 

    \addplot3[surf, shader = flat] 
    table[x index = 0, y index = 1, z index = 4] {\tableROM}; 

  \end{loglogaxis} 
\end{tikzpicture} %
  \tikzexternaldisable%
 
    \subcaption{Piecewise BT(20) approximation.}
  \end{subfigure}%
  \hfill%
  \begin{subfigure}[t]{.49\textwidth} 
    \centering
  \tikzexternalenable%
  \tikzsetnextfilename{piecewise_tbt_fixed_err_rel}%
  \filemodCmp{graphics/piecewise_tbt_fixed_err_rel.tikz}{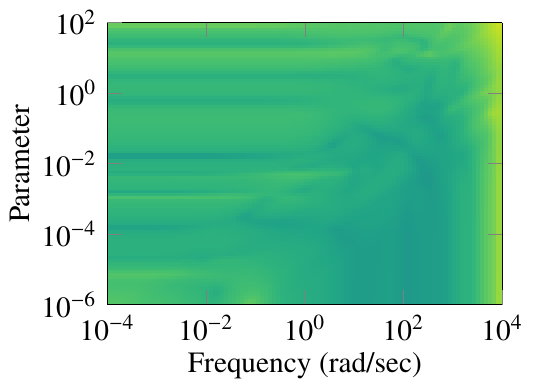}%
  {\tikzset{external/remake next}}{}%
  \begin{tikzpicture} 
  \pgfplotstableread{graphics/data/piecewise_BT_fixed_trom_all.dat}\tableROM 

  \begin{loglogaxis}[ 
    view={0}{90}, 
    width = .7\textwidth, 
    height = .5\textwidth, 
    scale only axis, 
    xmin = 1e-4, 
    xmax = 1e+4, 
    ymin = 1e-6, 
    ymax = 1e+2, 
    xtick = {1e-4, 1e-2, 1e0, 1e+2, 1e+4}, 
    ytick = {1e-6, 1e-4, 1e-2, 1e+0, 1e+2}, 
    zmode = log, 
    log base z = 10, 
    point meta min = -12, 
    point meta max = 4, 
    mesh/ordering = y varies, 
    mesh/rows = 100, 
    mesh/cols = 100, 
    xlabel = {\small Frequency (rad/sec)}, 
    xlabel style = {yshift = .3em}, 
    ylabel = {\small Parameter}, 
    ylabel style = {yshift = -.3em}, 
    scaled x ticks = false, 
    x tick label style = {/pgf/number format/fixed}] 

    \addplot3[surf, shader = flat] 
    table[x index = 0, y index = 1, z index = 4] {\tableROM}; 

  \end{loglogaxis} 
\end{tikzpicture} %
  \tikzexternaldisable%
 
    \subcaption{Truncated piecewise BT(20) approximation.}
  \end{subfigure} 

  \begin{subfigure}[t]{.49\textwidth} 
    \centering
  \tikzexternalenable%
  \tikzsetnextfilename{piecewise_irka_err_rel}%
  \filemodCmp{graphics/piecewise_irka_err_rel.tikz}{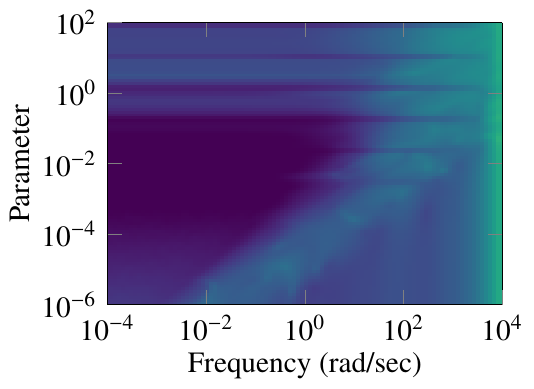}%
  {\tikzset{external/remake next}}{}%
  \begin{tikzpicture} 
  \pgfplotstableread{graphics/data/piecewise_irka_rom_all.dat}\tableROM 

  \begin{loglogaxis}[ 
    view={0}{90}, 
    width = .7\textwidth, 
    height = .5\textwidth, 
    scale only axis, 
    xmin = 1e-4, 
    xmax = 1e+4, 
    ymin = 1e-6, 
    ymax = 1e+2, 
    xtick = {1e-4, 1e-2, 1e0, 1e+2, 1e+4}, 
    ytick = {1e-6, 1e-4, 1e-2, 1e+0, 1e+2}, 
    zmode = log, 
    log base z = 10, 
    point meta min = -12, 
    point meta max = 4, 
    mesh/ordering = y varies, 
    mesh/rows = 100, 
    mesh/cols = 100, 
    xlabel = {\small Frequency (rad/sec)}, 
    xlabel style = {yshift = .3em}, 
    ylabel = {\small Parameter}, 
    ylabel style = {yshift = -.3em}, 
    scaled x ticks = false, 
    x tick label style = {/pgf/number format/fixed}] 

    \addplot3[surf, shader = flat] 
    table[x index = 0, y index = 1, z index = 4] {\tableROM}; 

  \end{loglogaxis} 
\end{tikzpicture} %
  \tikzexternaldisable%
 
    \subcaption{Piecewise IRKA approximation.} 
  \end{subfigure}%
  \hfill%
  \begin{subfigure}[t]{.49\textwidth} 
    \centering
  \tikzexternalenable%
  \tikzsetnextfilename{piecewise_tirka_err_rel}%
  \filemodCmp{graphics/piecewise_tirka_err_rel.tikz}{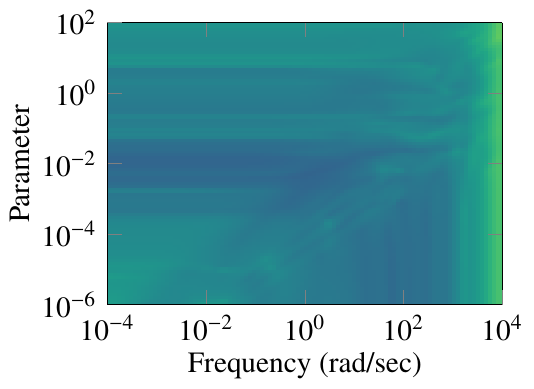}%
  {\tikzset{external/remake next}}{}%
  \begin{tikzpicture} 
  \pgfplotstableread{graphics/data/piecewise_irka_trom_all.dat}\tableROM 

  \begin{loglogaxis}[ 
    view={0}{90}, 
    width = .7\textwidth, 
    height = .5\textwidth, 
    scale only axis, 
    xmin = 1e-4, 
    xmax = 1e+4, 
    ymin = 1e-6, 
    ymax = 1e+2, 
    xtick = {1e-4, 1e-2, 1e0, 1e+2, 1e+4}, 
    ytick = {1e-6, 1e-4, 1e-2, 1e+0, 1e+2}, 
    zmode = log, 
    log base z = 10, 
    point meta min = -12, 
    point meta max = 4, 
    mesh/ordering = y varies, 
    mesh/rows = 100, 
    mesh/cols = 100, 
    xlabel = {\small Frequency (rad/sec)}, 
    xlabel style = {yshift = .3em}, 
    ylabel = {\small Parameter}, 
    ylabel style = {yshift = -.3em}, 
    scaled x ticks = false, 
    x tick label style = {/pgf/number format/fixed}] 

    \addplot3[surf, shader = flat] 
    table[x index = 0, y index = 1, z index = 4] {\tableROM}; 

  \end{loglogaxis} 
\end{tikzpicture} %
  \tikzexternaldisable%
 
    \subcaption{Truncated piecewise IRKA approximation.} 
  \end{subfigure} 
  \vspace{.5\baselineskip} 

  \tikzexternalenable%
  \tikzsetnextfilename{piecewise_cbar}%
  \filemodCmp{graphics/piecewise_cbar.tikz}{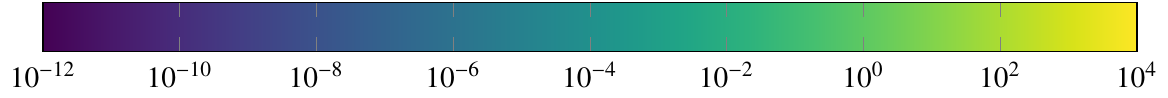}%
  {\tikzset{external/remake next}}{}%
  \begin{tikzpicture} 
  \begin{axis}[%
    hide axis, 
    scale only axis, 
    width = .95\textwidth,
    height = 1em, 
    point meta min = -12, 
    point meta max = 4, 
    colorbar, 
    colorbar horizontal, 
    colorbar style = { 
      xticklabel = $10^{\pgfmathparse{\tick} 
        \pgfmathprintnumber\pgfmathresult}$, 
      at = {(.5, 0)}, 
      anchor = north}, 
    scaled x ticks = false, 
    x tick label style = {/pgf/number format/fixed}] 
  \end{axis} 
\end{tikzpicture} %
  \tikzexternaldisable%

  \caption{Relative sigma-magnitude errors of different piecewise parametric
    reduction approaches for the thermal block model.}%
  \label{fig:BKS20-thermalblock_piecewise} 
\end{figure} 
\begin{figure}[tbp] 
  \centering
  \begin{subfigure}[t]{.49\textwidth} 
    \centering 
  \tikzexternalenable%
  \tikzsetnextfilename{piecewise_obt_tol_err_rel}%
  \filemodCmp{graphics/piecewise_obt_tol_err_rel.tikz}{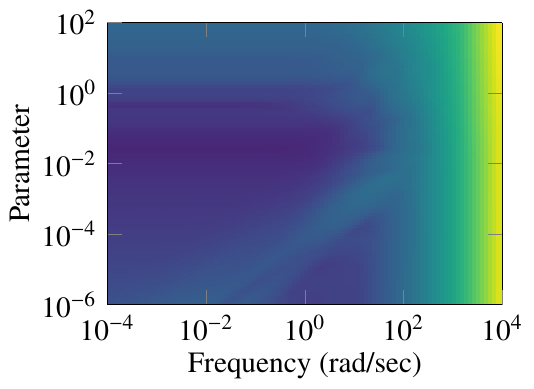}%
  {\tikzset{external/remake next}}{}%
  \begin{tikzpicture} 
  \pgfplotstableread{graphics/data/piecewise_BT_tol_orom_all.dat}\tableROM 

  \begin{loglogaxis}[ 
    view={0}{90}, 
    width = .7\textwidth, 
    height = .5\textwidth, 
    scale only axis, 
    xmin = 1e-4, 
    xmax = 1e+4, 
    ymin = 1e-6, 
    ymax = 1e+2, 
    xtick = {1e-4, 1e-2, 1e0, 1e+2, 1e+4}, 
    ytick = {1e-6, 1e-4, 1e-2, 1e+0, 1e+2}, 
    zmode = log, 
    log base z = 10, 
    point meta min = -12, 
    point meta max = 4, 
    mesh/ordering = y varies, 
    mesh/rows = 100, 
    mesh/cols = 100, 
    xlabel = {\small Frequency (rad/sec)}, 
    xlabel style = {yshift = .3em}, 
    ylabel = {\small Parameter}, 
    ylabel style = {yshift = -.3em}, 
    scaled x ticks = false, 
    x tick label style = {/pgf/number format/fixed}] 

    \addplot3[surf, shader = flat] 
    table[x index = 0, y index = 1, z index = 4] {\tableROM}; 

  \end{loglogaxis} 
\end{tikzpicture} %
  \tikzexternaldisable%
 
    \subcaption{Piecewise BT(\(10^{-4}\)) approximation.}
  \end{subfigure}%
  \hfill%
  \begin{subfigure}[t]{.49\textwidth} 
    \centering
  \tikzexternalenable%
  \tikzsetnextfilename{piecewise_tobt_tol_err_rel}%
  \filemodCmp{graphics/piecewise_tobt_tol_err_rel.tikz}{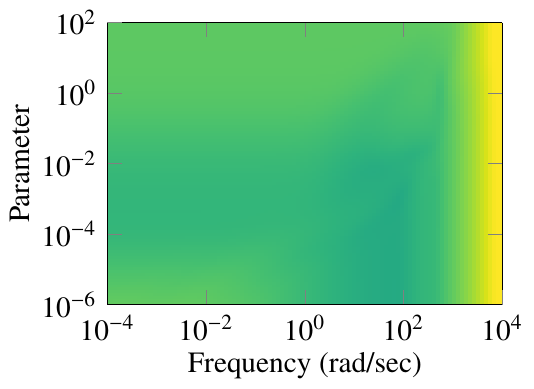}%
  {\tikzset{external/remake next}}{}%
  \begin{tikzpicture} 
  \pgfplotstableread{graphics/data/piecewise_BT_tol_torom_all.dat}\tableROM 

  \begin{loglogaxis}[ 
    view={0}{90}, 
    width = .7\textwidth, 
    height = .5\textwidth, 
    scale only axis, 
    xmin = 1e-4, 
    xmax = 1e+4, 
    ymin = 1e-6, 
    ymax = 1e+2, 
    xtick = {1e-4, 1e-2, 1e0, 1e+2, 1e+4}, 
    ytick = {1e-6, 1e-4, 1e-2, 1e+0, 1e+2}, 
    zmode = log, 
    log base z = 10, 
    point meta min = -12, 
    point meta max = 4, 
    mesh/ordering = y varies, 
    mesh/rows = 100, 
    mesh/cols = 100, 
    xlabel = {\small Frequency (rad/sec)}, 
    xlabel style = {yshift = .3em}, 
    ylabel = {\small Parameter}, 
    ylabel style = {yshift = -.3em}, 
    scaled x ticks = false, 
    x tick label style = {/pgf/number format/fixed}] 

    \addplot3[surf, shader = flat] 
    table[x index = 0, y index = 1, z index = 4] {\tableROM}; 

  \end{loglogaxis} 
\end{tikzpicture} %
  \tikzexternaldisable%
 
    \subcaption{Truncated piecewise BT(\(10^{-4}\)) approximation.}
  \end{subfigure} 

  \begin{subfigure}[t]{.49\textwidth} 
    \centering 
  \tikzexternalenable%
  \tikzsetnextfilename{piecewise_obt_fixed_err_rel}%
  \filemodCmp{graphics/piecewise_obt_fixed_err_rel.tikz}{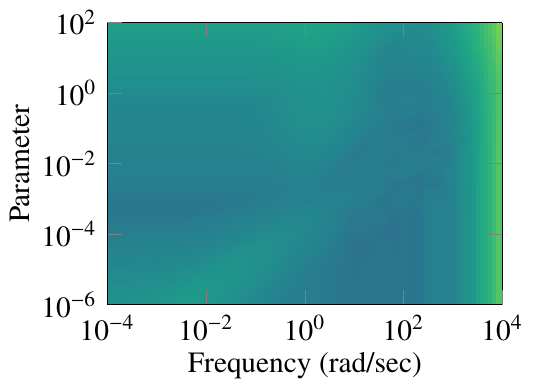}%
  {\tikzset{external/remake next}}{}%
  \begin{tikzpicture} 
  \pgfplotstableread{graphics/data/piecewise_BT_fixed_orom_all.dat}\tableROM 

  \begin{loglogaxis}[ 
    view={0}{90}, 
    width = .7\textwidth, 
    height = .5\textwidth, 
    scale only axis, 
    xmin = 1e-4, 
    xmax = 1e+4, 
    ymin = 1e-6, 
    ymax = 1e+2, 
    xtick = {1e-4, 1e-2, 1e0, 1e+2, 1e+4}, 
    ytick = {1e-6, 1e-4, 1e-2, 1e+0, 1e+2}, 
    zmode = log, 
    log base z = 10, 
    point meta min = -12, 
    point meta max = 4, 
    mesh/ordering = y varies, 
    mesh/rows = 100, 
    mesh/cols = 100, 
    xlabel = {\small Frequency (rad/sec)}, 
    xlabel style = {yshift = .3em}, 
    ylabel = {\small Parameter}, 
    ylabel style = {yshift = -.3em}, 
    scaled x ticks = false, 
    x tick label style = {/pgf/number format/fixed}] 

    \addplot3[surf, shader = flat] 
    table[x index = 0, y index = 1, z index = 4] {\tableROM}; 

  \end{loglogaxis} 
\end{tikzpicture} %
  \tikzexternaldisable%
 
    \subcaption{Piecewise BT(20) approximation.}
  \end{subfigure}%
  \hfill%
  \begin{subfigure}[t]{.49\textwidth} 
    \centering
  \tikzexternalenable%
  \tikzsetnextfilename{piecewise_tobt_fixed_err_rel}%
  \filemodCmp{graphics/piecewise_tobt_fixed_err_rel.tikz}{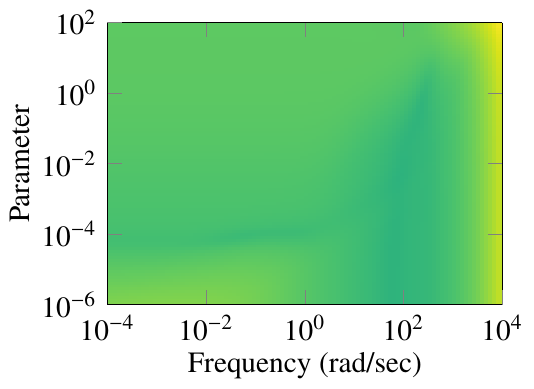}%
  {\tikzset{external/remake next}}{}%
  \begin{tikzpicture} 
  \pgfplotstableread{graphics/data/piecewise_BT_fixed_torom_all.dat}\tableROM 

  \begin{loglogaxis}[ 
    view={0}{90}, 
    width = .7\textwidth, 
    height = .5\textwidth, 
    scale only axis, 
    xmin = 1e-4, 
    xmax = 1e+4, 
    ymin = 1e-6, 
    ymax = 1e+2, 
    xtick = {1e-4, 1e-2, 1e0, 1e+2, 1e+4}, 
    ytick = {1e-6, 1e-4, 1e-2, 1e+0, 1e+2}, 
    zmode = log, 
    log base z = 10, 
    point meta min = -12, 
    point meta max = 4, 
    mesh/ordering = y varies, 
    mesh/rows = 100, 
    mesh/cols = 100, 
    xlabel = {\small Frequency (rad/sec)}, 
    xlabel style = {yshift = .3em}, 
    ylabel = {\small Parameter}, 
    ylabel style = {yshift = -.3em}, 
    scaled x ticks = false, 
    x tick label style = {/pgf/number format/fixed}] 

    \addplot3[surf, shader = flat] 
    table[x index = 0, y index = 1, z index = 4] {\tableROM}; 

  \end{loglogaxis} 
\end{tikzpicture} %
  \tikzexternaldisable%
 
    \subcaption{Truncated piecewise BT(20) approximation.}
  \end{subfigure} 

  \begin{subfigure}[t]{.49\textwidth} 
    \centering
  \tikzexternalenable%
  \tikzsetnextfilename{piecewise_oirka_err_rel}%
  \filemodCmp{graphics/piecewise_oirka_err_rel.tikz}{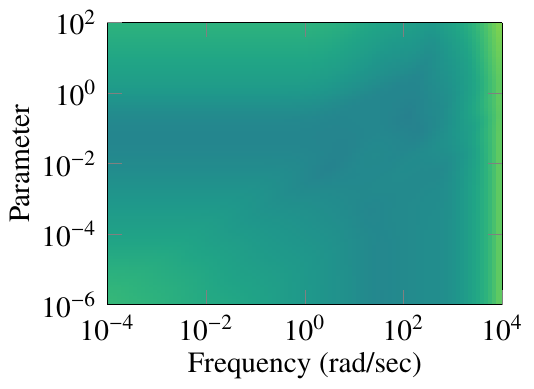}%
  {\tikzset{external/remake next}}{}%
  \begin{tikzpicture} 
  \pgfplotstableread{graphics/data/piecewise_irka_orom_all.dat}\tableROM 

  \begin{loglogaxis}[ 
    view={0}{90}, 
    width = .7\textwidth, 
    height = .5\textwidth, 
    scale only axis, 
    xmin = 1e-4, 
    xmax = 1e+4, 
    ymin = 1e-6, 
    ymax = 1e+2, 
    xtick = {1e-4, 1e-2, 1e0, 1e+2, 1e+4}, 
    ytick = {1e-6, 1e-4, 1e-2, 1e+0, 1e+2}, 
    zmode = log, 
    log base z = 10, 
    point meta min = -12, 
    point meta max = 4, 
    mesh/ordering = y varies, 
    mesh/rows = 100, 
    mesh/cols = 100, 
    xlabel = {\small Frequency (rad/sec)}, 
    xlabel style = {yshift = .3em}, 
    ylabel = {\small Parameter}, 
    ylabel style = {yshift = -.3em}, 
    scaled x ticks = false, 
    x tick label style = {/pgf/number format/fixed}] 

    \addplot3[surf, shader = flat] 
    table[x index = 0, y index = 1, z index = 4] {\tableROM}; 

  \end{loglogaxis} 
\end{tikzpicture} %
  \tikzexternaldisable%
 
    \subcaption{Piecewise IRKA approximation.} 
  \end{subfigure}%
  \hfill%
  \begin{subfigure}[t]{.49\textwidth} 
    \centering
  \tikzexternalenable%
  \tikzsetnextfilename{piecewise_toirka_err_rel}%
  \filemodCmp{graphics/piecewise_toirka_err_rel.tikz}{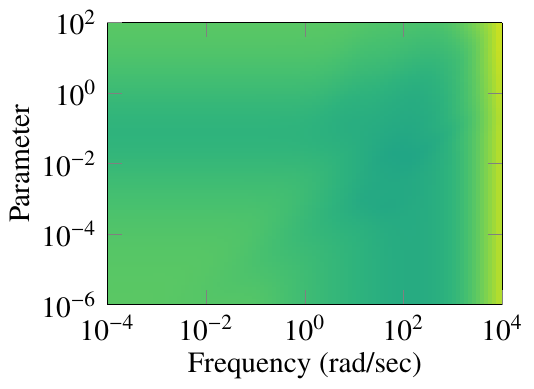}%
  {\tikzset{external/remake next}}{}%
  \begin{tikzpicture} 
  \pgfplotstableread{graphics/data/piecewise_irka_torom_all.dat}\tableROM 

  \begin{loglogaxis}[ 
    view={0}{90}, 
    width = .7\textwidth, 
    height = .5\textwidth, 
    scale only axis, 
    xmin = 1e-4, 
    xmax = 1e+4, 
    ymin = 1e-6, 
    ymax = 1e+2, 
    xtick = {1e-4, 1e-2, 1e0, 1e+2, 1e+4}, 
    ytick = {1e-6, 1e-4, 1e-2, 1e+0, 1e+2}, 
    zmode = log, 
    log base z = 10, 
    point meta min = -12, 
    point meta max = 4, 
    mesh/ordering = y varies, 
    mesh/rows = 100, 
    mesh/cols = 100, 
    xlabel = {\small Frequency (rad/sec)}, 
    xlabel style = {yshift = .3em}, 
    ylabel = {\small Parameter}, 
    ylabel style = {yshift = -.3em}, 
    scaled x ticks = false, 
    x tick label style = {/pgf/number format/fixed}] 

    \addplot3[surf, shader = flat] 
    table[x index = 0, y index = 1, z index = 4] {\tableROM}; 

  \end{loglogaxis} 
\end{tikzpicture} %
  \tikzexternaldisable%
 
    \subcaption{Truncated piecewise IRKA approximation.} 
  \end{subfigure} 
  \vspace{.5\baselineskip} 

  \tikzexternalenable%
  \tikzsetnextfilename{piecewise_cbar}%
  \filemodCmp{graphics/piecewise_cbar.tikz}{graphics/externalize/piecewise_cbar.pdf}%
  {\tikzset{external/remake next}}{}%
  \begin{tikzpicture} 
  \begin{axis}[%
    hide axis, 
    scale only axis, 
    width = .95\textwidth,
    height = 1em, 
    point meta min = -12, 
    point meta max = 4, 
    colorbar, 
    colorbar horizontal, 
    colorbar style = { 
      xticklabel = $10^{\pgfmathparse{\tick} 
        \pgfmathprintnumber\pgfmathresult}$, 
      at = {(.5, 0)}, 
      anchor = north}, 
    scaled x ticks = false, 
    x tick label style = {/pgf/number format/fixed}] 
  \end{axis} 
\end{tikzpicture} %
  \tikzexternaldisable%

  \caption{Relative sigma-magnitude errors of different piecewise parametric
    one-sided reduction approaches for the thermal block model.}%
  \label{fig:BKS20-thermalblock_onesided_piecewise} 
\end{figure} 
\begin{figure}[tbp]
  \centering
  \begin{subfigure}[t]{.49\textwidth} 
    \centering 
  \tikzexternalenable%
  \tikzsetnextfilename{interp_bt_err_rel}%
  \filemodCmp{graphics/interp_bt_err_rel.tikz}{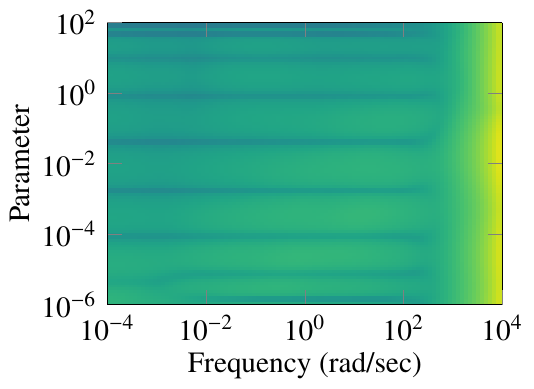}%
  {\tikzset{external/remake next}}{}%
  \begin{tikzpicture} 
  \pgfplotstableread{graphics/data/interp_BT_rom_all.dat}\tableROM 

  \begin{loglogaxis}[ 
    view={0}{90}, 
    width = .7\textwidth, 
    height = .5\textwidth, 
    scale only axis, 
    xmin = 1e-4, 
    xmax = 1e+4, 
    ymin = 1e-6, 
    ymax = 1e+2, 
    xtick = {1e-4, 1e-2, 1e0, 1e+2, 1e+4}, 
    ytick = {1e-6, 1e-4, 1e-2, 1e+0, 1e+2}, 
    zmode = log, 
    log base z = 10, 
    point meta min = -9, 
    point meta max = 4, 
    mesh/ordering = y varies, 
    mesh/rows = 100, 
    mesh/cols = 100, 
    xlabel = {\small Frequency (rad/sec)}, 
    xlabel style = {yshift = .3em}, 
    ylabel = {\small Parameter}, 
    ylabel style = {yshift = -.3em}, 
    scaled x ticks = false, 
    x tick label style = {/pgf/number format/fixed}] 

    \addplot3[surf, shader = flat] 
    table[x index = 0, y index = 1, z index = 4] {\tableROM}; 

  \end{loglogaxis} 
\end{tikzpicture} %
  \tikzexternaldisable%
 
    \subcaption{Lagrange-BT approximation.} 
  \end{subfigure}%
  \hfill%
  \begin{subfigure}[t]{.49\textwidth} 
    \centering
  \tikzexternalenable%
  \tikzsetnextfilename{bspline_bt_err_rel}%
  \filemodCmp{graphics/bspline_bt_err_rel.tikz}{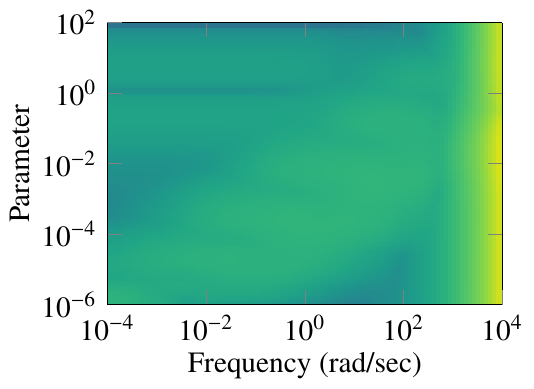}%
  {\tikzset{external/remake next}}{}%
  \begin{tikzpicture} 
  \pgfplotstableread{graphics/data/bspline_BT_rom_all.dat}\tableROM 

  \begin{loglogaxis}[ 
    view={0}{90}, 
    width = .7\textwidth, 
    height = .5\textwidth, 
    scale only axis, 
    xmin = 1e-4, 
    xmax = 1e+4, 
    ymin = 1e-6, 
    ymax = 1e+2, 
    xtick = {1e-4, 1e-2, 1e0, 1e+2, 1e+4}, 
    ytick = {1e-6, 1e-4, 1e-2, 1e+0, 1e+2}, 
    zmode = log, 
    log base z = 10, 
    point meta min = -9, 
    point meta max = 4, 
    mesh/ordering = y varies, 
    mesh/rows = 100, 
    mesh/cols = 100, 
    xlabel = {\small Frequency (rad/sec)}, 
    xlabel style = {yshift = .3em}, 
    ylabel = {\small Parameter}, 
    ylabel style = {yshift = -.3em}, 
    scaled x ticks = false, 
    x tick label style = {/pgf/number format/fixed}] 

    \addplot3[surf, shader = flat] 
    table[x index = 0, y index = 1, z index = 4] {\tableROM}; 

  \end{loglogaxis} 
\end{tikzpicture} %
  \tikzexternaldisable%
 
    \subcaption{Bspline-BT approximation.} 
  \end{subfigure} 

  \begin{subfigure}[t]{.49\textwidth} 
    \centering
  \tikzexternalenable%
  \tikzsetnextfilename{interp_irka_err_rel}%
  \filemodCmp{graphics/interp_irka_err_rel.tikz}{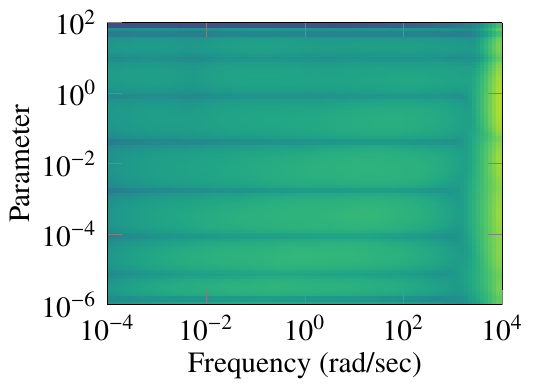}%
  {\tikzset{external/remake next}}{}%
  \begin{tikzpicture} 
  \pgfplotstableread{graphics/data/interp_IRKA_rom_all.dat}\tableROM 

  \begin{loglogaxis}[ 
    view={0}{90}, 
    width = .7\textwidth, 
    height = .5\textwidth, 
    scale only axis, 
    xmin = 1e-4, 
    xmax = 1e+4, 
    ymin = 1e-6, 
    ymax = 1e+2, 
    xtick = {1e-4, 1e-2, 1e0, 1e+2, 1e+4}, 
    ytick = {1e-6, 1e-4, 1e-2, 1e+0, 1e+2}, 
    zmode = log, 
    log base z = 10, 
    point meta min = -9, 
    point meta max = 4, 
    mesh/ordering = y varies, 
    mesh/rows = 100, 
    mesh/cols = 100, 
    xlabel = {\small Frequency (rad/sec)}, 
    xlabel style = {yshift = .3em}, 
    ylabel = {\small Parameter}, 
    ylabel style = {yshift = -.3em}, 
    scaled x ticks = false, 
    x tick label style = {/pgf/number format/fixed}] 

    \addplot3[surf, shader = flat] 
    table[x index = 0, y index = 1, z index = 4] {\tableROM}; 

  \end{loglogaxis} 
\end{tikzpicture} %
  \tikzexternaldisable%
 
    \subcaption{Lagrange-IRKA approximation.} 
  \end{subfigure}%
  \hfill%
  \begin{subfigure}[t]{.49\textwidth} 
    \centering
  \tikzexternalenable%
  \tikzsetnextfilename{bspline_irka_err_rel}%
  \filemodCmp{graphics/bspline_irka_err_rel.tikz}{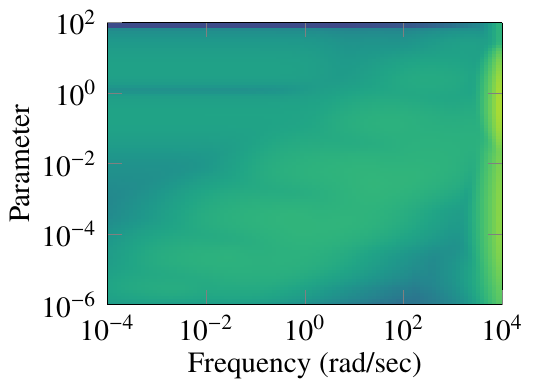}%
  {\tikzset{external/remake next}}{}%
  \begin{tikzpicture} 
  \pgfplotstableread{graphics/data/bspline_IRKA_rom_all.dat}\tableROM 

  \begin{loglogaxis}[ 
    view={0}{90}, 
    width = .7\textwidth, 
    height = .5\textwidth, 
    scale only axis, 
    xmin = 1e-4, 
    xmax = 1e+4, 
    ymin = 1e-6, 
    ymax = 1e+2, 
    xtick = {1e-4, 1e-2, 1e0, 1e+2, 1e+4}, 
    ytick = {1e-6, 1e-4, 1e-2, 1e+0, 1e+2}, 
    zmode = log, 
    log base z = 10, 
    point meta min = -9, 
    point meta max = 4, 
    mesh/ordering = y varies, 
    mesh/rows = 100, 
    mesh/cols = 100, 
    xlabel = {\small Frequency (rad/sec)}, 
    xlabel style = {yshift = .3em}, 
    ylabel = {\small Parameter}, 
    ylabel style = {yshift = -.3em}, 
    scaled x ticks = false, 
    x tick label style = {/pgf/number format/fixed}] 

    \addplot3[surf, shader = flat] 
    table[x index = 0, y index = 1, z index = 4] {\tableROM}; 

  \end{loglogaxis} 
\end{tikzpicture} %
  \tikzexternaldisable%
 
    \subcaption{Bspline-IRKA approximation.} 
  \end{subfigure} 
  \vspace{.5\baselineskip} 

  \tikzexternalenable%
  \tikzsetnextfilename{interp_cbar}%
  \filemodCmp{graphics/interp_cbar.tikz}{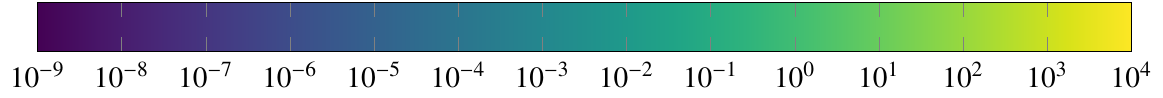}%
  {\tikzset{external/remake next}}{}%
  \begin{tikzpicture} 
  \begin{axis}[%
    hide axis, 
    scale only axis, 
    width = .95\textwidth,
    height = 1em, 
    point meta min = -9, 
    point meta max = 4, 
    colorbar, 
    colorbar horizontal, 
    colorbar style = { 
      xticklabel = $10^{\pgfmathparse{\tick} 
        \pgfmathprintnumber\pgfmathresult}$, 
      at = {(.5, 0)}, 
      anchor = north}, 
    scaled x ticks = false, 
    x tick label style = {/pgf/number format/fixed}] 
  \end{axis} 
\end{tikzpicture} %
  \tikzexternaldisable%

  \caption{Relative sigma-magnitude errors of different transfer function
    interpolation methods for parametric reduction for the thermal block
    model.}%
  \label{fig:BKS20-thermalblock_interpolation} 
\end{figure}
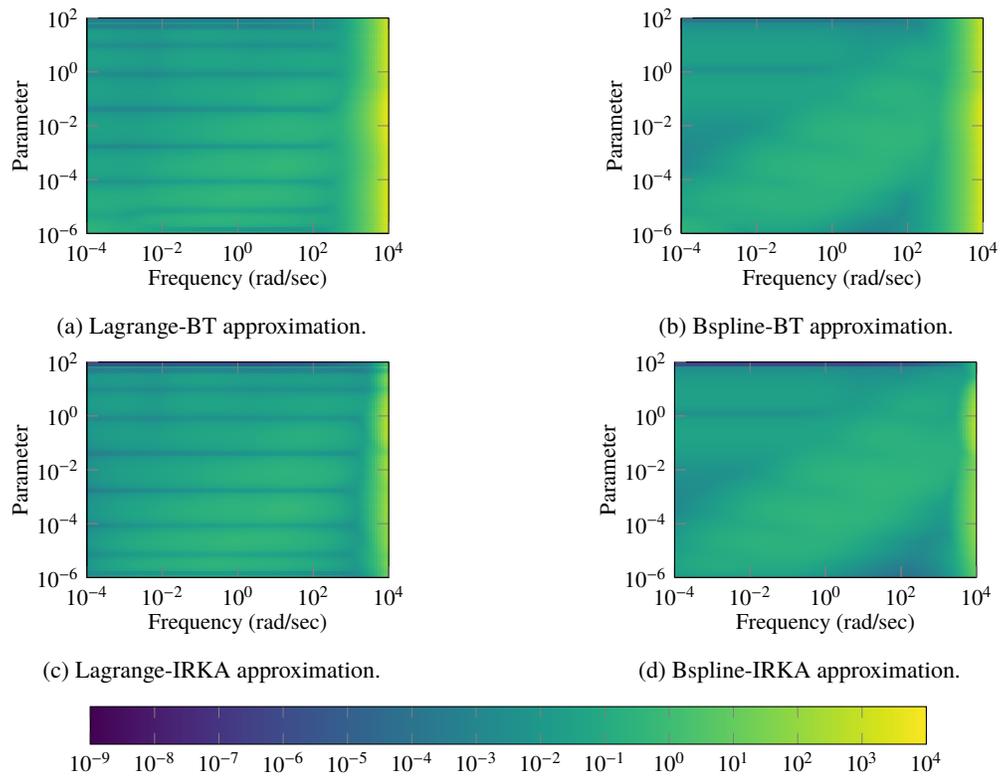 
\begin{table}\centering
  \begin{tabular}{|l|c|c|c|}
    \hline
    Method &ROMs &Full& One-sided\\\hline
    \multicolumn{3}{c}{\bfseries Piecewise}\\\hline
    BT(\(10^{-4}\))&9/12/15/13/12/9/8/9/8/7&102~(52)&200 (36)\\\hline 
    BT(20)        &20/20/20/20/20/20/20/20/20/20&199 (64)&200 (72)\\\hline 
    IRKA          &20/20/20/20/20/20/20/20/20/20&200 (132)&200 (132)\\\hline
    \multicolumn{3}{c}{\bfseries Lagrange}\\\hline
    BT(\(10^{-4}\))&9/9/12/15/12/9/8/8/7/7&96&--\\\hline 
    IRKA          &20/20/20/20/20/20/20/20/20/20&200&--\\\hline
    \multicolumn{3}{c}{\bfseries B-Spline}\\\hline
    BT(\(10^{-4}\))&9/9/12/15/12/9/8/8/7/7&96&--\\\hline 
    IRKA          &20/20/20/20/20/20/20/20/20/20&200&--\\\hline
  \end{tabular}
  \caption{Reduced orders of the training-sample ROMs and final
    ROM\ (numbers in () are after additional truncation with tolerance
    \(10^{-6}\)).}%
  \label{tab:BKS20-ROM-orders}
\end{table}
The experiments reported here have been executed in \matlab{} R2019a on a Lenovo
X380 Yoga equipped with an \intel{} i7 8770 and 32GB of main memory running
64bit Linux based on Ubuntu 18.04. The experiments use
\mmess{}-2.0.1~\cite{SaaKB-mmess-all-versions} and Chebfun version
5.7.0~\cite{DriHT14,ChebfunWeb}.
\begin{center}
  \fbox{%
    \parbox{.9\linewidth}{
      The source code of the implementations used to compute the presented
      results can be obtained from:
      \begin{center} 
        \url{https://doi.org/10.5281/zenodo.3678213}
      \end{center}
      and is authored by Jens Saak and Steffen W.~R. Werner.
    }%
  }%
\end{center}

For easier comparison with the other reported software packages, all
experiments use the thermal block benchmark introduced in a separate chapter of
this volume. It describes a simple heat transfer model on the domain depicted in
Figure~\ref{fig:BKS20-FOM-domain}. Here, we investigate the one parameter
version of the benchmark. That means, the heat transfer coefficients on the four
circular sub-domains are given as \(0.2\mu\), \(0.4\mu\), \(0.6\mu\), and
\(0.8\mu\) for a single scalar parameter
\(\mu\in[10^{-6},10^{2}]=M\subset\R\). The full order model has dimension
\(n=7\,488\) and one input but 4 outputs. In Figure~\ref{fig:BKS20-FOM-sigma} we
present the sigma-magnitude plot of the full order model (FOM), i.e.\ we plot
\(\norm[2]{H(\mu,s)}=\sigma_{\max}(H(\mu,s))\) over the full parameter range and
the frequency range \([10^{-4},10^{4}]\). The plot is based on 100
logarithmically equi-spaced sample points (\texttt{logspace}-generated) in each
direction. We also use this sampling for all relative sigma-magnitude error
plots in the other figures. The error plots analogously show
\(\norm[2]{H(\mu,s)-\hH(\mu,s)} / \norm[2]{H(\mu,s)}\).

Excluding the 10\,000 evaluations for the pre-sampling of the original transfer
function, all computations for generation of the ROMs and
evaluation of the approximation errors can be executed in less than 8 minutes.

We compare both IRKA and classic (Lyapunov) balanced truncation (BT) in the
piecewise as well as the transfer function interpolation context. For IRKA we
use fixed order \(r=20\) in all training samples, while for BT we run in two
modes. Since we have the BT error-bound that allows for adaptive processing,
i.e.\ automatic choice of the reduced order, we do that with absolute error
tolerance \(10^{-4}\). On the other hand, for a more fair comparison to IRKA we
also run BT for fixed order \(r=20\). We refer to these two modes as
\(\text{BT}(10^{-4})\) and \(\text{BT}(20)\) in the following.

For the piecewise approaches we use 10 logarithmically equi-spaced
(\texttt{logspace}-generated) parameter samples in \(M\) as the training
positions. For the interpolatory approaches we choose 10 Chebyshev-roots
generated by Chebfun. We have mentioned the final rank-truncation after basis
concatenation in Section~\ref{sec:BKS20-piecewise-mor}. We use a tolerance equal
to \texttt{eps} in the standard case. Alternatively, to further compress the
final parametric ROM, we truncate with tolerance \(10^{-6}\) and refer to this
approach by the name {\em truncated piecewise}.

For the training, BT can not reuse information from previous samples very
easily. On the other hand, IRKA can be initialized with the ROM from the
previous parameter sample, which in most cases made it converge after less than
5 steps (mostly being stopped by criterion monitoring the relative change of
the model in the \(\cH_{2}\)-norm). For further implementation details we refer
to the scripts in the code package.

Although, BT guarantees the local ROMs in the sample points to preserve the
asymptotic stability of the original model, and also IRKA preserves stability
upon convergence, this feature is in general lost after concatenating the bases
to the global one. Still, for a one-sided projection the stability of the global
ROM can be preserved. Due to stability and symmetry of the thermal block model,
Bendixson's theorem~\cite{Ben02a} guarantees this. Therefore, we compare to a
one-sided  approach that 
simply combines \(V\) and \(W\) into one matrix. The comparison can be found in
Figures~\ref{fig:BKS20-thermalblock_piecewise}
and~\ref{fig:BKS20-thermalblock_onesided_piecewise}. And the corresponding ROM
orders are given in the first block of Table~\ref{tab:BKS20-ROM-orders}.

For the interpolatory approaches, we compare Lagrange polynomials and variation
diminishing B-splines of order 2. Here, we always use \(\text{BT}(10^{-4})\) in
the BT case, since the results are already hard to distinguish from the
IRKA-based ones in this case and we do not expect much improvement from the
higher local orders. 

It can be seen from Table~\ref{tab:BKS20-ROM-orders} that the piecewise BT
models are, in parts significantly, smaller than the piecewise IRKA models. This
comes at the price that the accuracy is not as good in parts of the
domain. Nonetheless, e.g.\ the truncated one-sided \(\text{BT}(10^{-4})\)
approximation yields a relative error of below 1\% on a majority (around 70\%)
of the investigated frequency-parameter domain with a model size that is \(3.7\)
to \(5.6\) times smaller. There is a significant increase in error for those
frequencies, where the transfer function has very small values (see
Figure~\ref{fig:BKS20-FOM-sigma}) that can be considered to be on the noise
level.

The results are very satisfactory and so are the computation times. This
indicates that the implementations can be used for larger and more challenging
examples, that we can not report here due to space restrictions. 

\bibliographystyle{plain}
\bibliography{mor,csc,software,mess-2.0}
\end{document}
